\documentclass[12pt]{iopart}

\usepackage[dvips]{graphicx}
\usepackage[usenames]{color}

\newcommand{\bm}[1]{\hbox{\boldmath{$#1$}}}
\newcommand{\sbm}[1]{\hbox{\boldmath{\scriptsize$#1$}}}
\newcommand{\Mp}{M_{\rm pl}}
\newcommand{\dd}{{\rm d}}
\newcommand{\gR}{{^g\!R}}
\newcommand{\sR}{{^s\!R}}

\newcommand{\gz}{{^g\!\zeta}}

\newcommand{\Approx}{\stackrel{\rm  IR}{\approx}}
\newcommand{\Gbz}{{^g\!\bar{\zeta}}}

\newcommand{\cv}{\{\zeta,\pi\}}
\newcommand{\cvt}{\{\tilde{\zeta},\tilde{\pi}\}}
\newcommand{\IR}{{\rm IR}}

\newcommand{\bfk}{{\mbox{\boldmath$k$}}}

\newcommand{\gauge}{{\em gauge} }
\newcommand{\cO}{{\cal O}}
\newcommand{\alt}{\hspace{0.3em}\raisebox{0.4ex}{$<$}\hspace{-0.75em}\raisebox{-.7ex}{$\sim$}\hspace{0.3em}}
\newcommand{\agt}{\hspace{0.3em}\raisebox{0.4ex}{$>$}\hspace{-0.75em}\raisebox{-.7ex}{$\sim$}\hspace{0.3em}}

\begin{document}

\title[Loops in inflationary correlation functions.]{Loops in inflationary
correlation functions.}

\author{Takahiro Tanaka$^{1}$, \, Yuko Urakawa$^{2}$}
\address{$^{1}$ Yukawa Institute for Theoretical Physics, Kyoto university,
  Kyoto, 606-8502, Japan\\
$^{2}$ Departament de F{\'\i}sica Fonamental i Institut de Ci{\`e}ncies del Cosmos, 
Universitat de Barcelona,
Mart{\'\i}\ i Franqu{\`e}s 1, 08028 Barcelona, Spain}
\eads{\mailto{tanaka@yukawa.kyoto-u.ac.jp},
\mailto{yurakawa@ffn.ub.es}}

\begin{abstract}
We review the recent progress regarding the loop 
corrections to the correlation functions in the inflationary 
universe. A naive 
perturbation theory predicts that loop corrections generated during 
inflation suffer from various infrared (IR) pathologies. 
Introducing an IR cutoff by hand is neither satisfactory nor enough to fix the problem of secular growth, 
which may ruin the predictive power of inflation models 
if the inflation lasts sufficiently long. 
We discuss the origin of the IR divergences and explore the regularity
conditions of loop corrections for the adiabatic perturbation, the iso-curvature perturbation, 
and the tensor perturbation, in turn. 
These three kinds of perturbations have qualitative 
differences, but in discussing the IR regularity 
there is a feature common to all cases,  
which is the importance of the proper 
identification of observable quantities. 
Genuinely observable quantities should respect the gauge invariance from 
the view point of a local observer. 
Interestingly, we find that the requirement of the IR regularity 
restricts the allowed quantum states.    
\end{abstract}

\maketitle

\section{Motivations and overview}
The Planck satellite has provided the most precise map of 
perturbation at around the recombination epoch of the universe 
through the measurement of the cosmic microwave background (CMB). 
The observed spectrum of the CMB strongly suggests that there was
an inflationary phase in the very early stage of the universe, 
which relaxes the fine tuning of the
initial condition of the Big Bang universe. 
When
we postulate the presence of a light scalar field, the inflaton, 
its quantum fluctuation is amplified to the observable level 
due to the rapid expansion during inflation, and it provides the
source of the large scale structure of the universe. 
Thus generated primordial perturbation becomes almost scale
invariant, which is consistent with the measurement of the CMB.  
 
Since the inflation paradigm is getting a more and more convincing
scenario of the early universe, it will be high time to re-examine all
the predictions based on this paradigm more carefully. 
In general relativity, non-linearity necessarily enters into the 
evolution of the primordial perturbation. 
Therefore, to compute the primordial perturbation, 
we need to understand interacting quantum fields in the inflationary 
universe. 
The non-linear quantum evolution was initially discussed by
considering interacting scalar fields on a fixed background 
neglecting the gravitational fluctuation~\cite{BD}. 
The systematic study of 
quantum evolution of perturbation including the gravitational 
non-linearity in the context of realistic inflation models 
was initiated by Maldacena in
Ref.~\cite{Maldacena2002} more than 20 years after Bardeen's 
gauge invariant linear perturbation theory~\cite{Bardeen:1980kt}. 
As a consequence of the
general covariance, the description of gravitating system 
has the freedom in choosing the coordinates. 
To provide a theoretical prediction of the observable
fluctuation, we need to identify the gauge-invariant
degrees of freedom, excluding the gauge ambiguity.
Using the Arnowitt-Deser-Misner (ADM) formalism,
Maldacena derived the third-order action expressed only in terms of the
so-called gauge-invariant variable. 
Using this non-linear action, he also computed
the bi-spectrum of the primordial fluctuation at the tree level. The
non-Gaussianity is now within the reach of observations. 
The constraint on the non-Gaussianities given by the Planck satellite has 
already excluded a number of inflation models, 
which highlights the importance of studying the non-linear evolution of the
primordial fluctuation.

The non-linear evolution also generates loop corrections. The amplitude
of the loop corrections is typically suppressed by an extra power of the
amplitude of the power spectrum $\sim (H/\Mp)^2$, 
where $H$ denotes the Hubble parameter during inflation 
and $\Mp$ is the reduced Planck mass, defined by $\Mp^{-2}\equiv 8 \pi G$. 
However, the suppression by the factor $(H/\Mp)^2$
might be compensated by the accumulation of the infrared (IR)
contributions. 
When we assume the scale invariant spectrum in the IR limit, 
a naive loop integral results in a factor  
$\propto \int \dd^3 \bm{k}/k^3$, which 
leads to the IR divergence. Even if we introduce an IR cutoff, 
say at the Hubble scale at a time $t=t_0$,
integrating IR modes leads to the logarithmic secular growth as
\begin{eqnarray}
  \int^{a(t)H(t)}_{a(t_0) H(t_0)} \frac{\dd k}{k}
\sim \ln \frac{a(t)}{a(t_0)}\,, \label{Eq:secular}
\end{eqnarray}
where $a(t)$ denotes the scale factor. 
Therefore, the loop corrections may dominate
in case inflation continues for a sufficiently long period, 
leading to the breakdown of perturbative expansion.

This IR singular behavior due to the accumulated IR contributions 
associated with a massless scalar field 
 has long been known in de Sitter space~\cite{Ford:1977in,Allen:1987tz,Kirsten:1993ug}.
The study on the implication of this singular behaviour to interacting fields 
dates back to Refs.~\cite{Tagirov:1972vv, Sasaki:1992ux, Tsamis:1993ub} 
and afterwards many works such as Refs.~\cite{TW96, Onemli:2002hr,
Onemli:2004mb, Kahya:2006hc, Boyanovsky:2004gq, Brunier:2004sb, Boyanovsky:2004ph, Boyanovsky:2005sh,
Boyanovsky:2005px, Marozzi:2012tp} followed. A similar divergence has been reported for
various fields~\cite{Prokopec:2002jn, Miao:2005am, Prokopec:2006ue,
Prokopec:2007ak}. The accumulated effect of the IR contributions has
been explored also in the presence of the gravitational
fluctuations~\cite{TW962, ABM, Sloth:2006az, Sloth:2006nu, Seery:2007we,
Seery:2007wf, Urakawa:2008rb, Cogollo:2008bi, Rodriguez:2008hy,
Bartolo:2010bu, Kitamoto:2012vj, Kitamoto:2013rea, Leonard:2012fs, Leonard:2013xsa,
	Kahya:2010xh, Marozzi:2013uva}. 
The IR contributions for multi scalar
fields also have been studied~\cite{Adshead:2008gk, Gao:2009fx, RB}. 
(A thorough overview of the 
historical progress regarding the IR issues is nicely summarized in the review paper by
Seery in Ref.~\cite{Seery:2010kh}.) 
The accumulation of the IR modes may cause 
the secular modification of the effective theory. 
Tsamis and Woodard claimed that the IR gravitons may
screen the cosmological constant, 
which may explain the unnaturally small observed cosmological constant (the
cosmological constant problem)~\cite{TW962}. 
The possible secular effects
have been further examined by Woodard
{\it et al.}~\cite{TW96, Onemli:2002hr, Onemli:2004mb} and more recently by
Polyakov {\it et al.} \cite{Polyakov:2007mm, Polyakov:2009nq,
Krotov:2010ma}, 
by Alvarez and Vidal \cite{Alvarez:2010te}, and also by
Kitamoto and Kitazawa~\cite{Kitamoto:2012vj, Kitamoto:2013rea, KK10, KK1012, KK11}.  
 (See also the lecture note \cite{Romania:2012av} and the references therein.) 
With these observations, one may worry if 
we cannot provide sound theoretical predictions based on the 
inflationary universe paradigm. 
However, it is not manifest whether the reported IR secular evolution
is a truly physical effect or not. In particular, 
once we include the fluctuation of the gravitational field, we need to
be careful in the discrimination between gauge-invariant 
physical degrees of freedom and spurious gauge artifacts.

Given that the
inflationary universe includes multi scalar fields, the physical 
degrees of freedom therein are decomposed into the three categories: 
the adiabatic perturbation, the iso-curvature
perturbation, and the tensor perturbation. The adiabatic
perturbation describes the fluctuation of the inflaton 
and the iso-curvature perturbation describes the remaining 
scalar degrees of freedom. Note that an appropriate description of 
the adiabatic perturbation and the tensor perturbation 
requires taking into account the gravitational perturbation. 
While the iso-curvature perturbation is primarily 
well-approximated by a test field in a fixed background. 
In this review our main focus is on the adiabatic perturbation, but 
we also discuss the other types of perturbation briefly.

The IR issues have been addressed based on various approaches and
approximations. The methodological variation may have made
the mutual relation among them obscure. 
In this review paper, we aim at giving a consistent 
explanation more than giving an exhaustive review on this subject. The
outline of this paper is as follows. In Sec.~\ref{Sec:preparation},
after we overview Maldacena's method to compute the non-linear
contributions, in Sec.~\ref{SSec:Vdiv} we will classify various
divergences which appear in computing loop corrections. The
appearance of IR divergences is not peculiar to the cosmological 
perturbation theory. In Sec.~\ref{SSec:QEDQCD}, we briefly overview the IR
divergences in QED and QCD and compare them to those in the
inflationary universe. In Sec.~\ref{Sec:GI}, we will show that
the singular behaviour of the IR contributions of the curvature
perturbation $\zeta$ is deeply related to the influence from the
outside of the observable universe, which can be rephrased as the gauge 
degrees of freedom in the local observable universe. In Sec.~\ref{Sec:SRV}, we will
present how to introduce observable quantities that 
preserve the gauge invariance.  Afterwards, we will
also show that requesting the regularity of the IR contributions is
equivalent to requesting the gauge invariance in the local universe. 
In Sec.~\ref{Sec:Euclidean}, we show that the Euclidean vacuum 
preserves the gauge invariance in the local universe and guarantees 
the regularity of the loop corrections for the curvature perturbation
$\zeta$. In Sec.~\ref{Sec:test}, we will summarize the recent progress
regarding the IR issues for a test field in a fixed inflationary
background, which is supposed to give a good approximation to the
iso-curvature perturbation. In Sec.~\ref{Sec:GW}, we will briefly discuss the IR
issues of the graviton loops. Finally, in Sec.~\ref{Sec:summary} we will summarize the
current status of this subject and will discuss the future issues.

In this review article, we will introduce the following abbreviations\\[2pt]
\begin{tabular}{lll}
 &superH : super Hubble & subH : sub Hubble \\
 &IR : infrared& UV : ultraviolet \\
 &tIR : transient IR (Sec.~\ref{IRtypeII}) & IRdiv : IR
     divergence (Sec.~\ref{IRtypeI}) \\
  &IRsec : IR secular growth (Sec.~\ref{IRtypeII}) &  \\
\end{tabular}
\\
\begin{tabular}{ll}
 & SG : secular growth due to temporal integral (Sec.~\ref{IRtypeIV})  \\
 & \gauge: coordinate choice in the local universe (Sec.~\ref{SSec:RDF})
\end{tabular}
\\[2pt]
where, in parentheses, we described the section in which the
abbreviation is introduced.  

\section{Issues in calculating loop corrections}   
\label{Sec:preparation}
In this review, we mainly focus on a single scalar field model. For
simplicity, we consider a scalar field with the canonical kinetic
term, whose action takes the form
\begin{eqnarray}
 S = \frac{1}{2} \int \sqrt{-g}~ [ M_{\rm pl}^2 R -
  g^{\mu\nu}\Phi_{,\mu}\Phi_{,\nu} - 2 U(\Phi) ] \dd^4x~. 
\end{eqnarray} 
An extension to a non-canonical kinetic term
is straightforward. 
In this section we explain our main concern 
in calculating the loop corrections to the correlation functions 
in this simple model. For later use, we rescale the variables as 
\begin{eqnarray}
 \phi \equiv  \Phi/ \Mp\,,\quad
 V(\phi) \equiv  U(\Phi)/ \Mp^2\,, 
\end{eqnarray}
so that $M_{\rm pl}^2$ is factored out from the action as 
\begin{eqnarray}
 S = \frac{M_{\rm pl}^2}{2} \int \sqrt{-g}~ [R - g^{\mu\nu}\phi_{,\mu} \phi_{,\nu} 
   - 2 V(\phi) ]\dd^4x~. \label{Exp:S}
\end{eqnarray}
Then, the equations of
motion and the constraint equations do not depend on the Planck mass.  

\subsection{The action with non-locality}
The ADM formalism is well suited 
for the non-linear analysis of the Einstein gravity~\cite{Maldacena2002, Seery:2005wm}. 
The ADM line element is expressed  as  
\begin{eqnarray}
 \dd s^2 = - N^2 \dd t^2  + h_{ij} (\dd x^i + N^i \dd t) (\dd x^j + N^j \dd
  t)~,
 \label{Exp:ADMmetric}
\end{eqnarray}
where we introduced the lapse function $N$, the
shift vector $N^i$, and the purely spatial metric $h_{ij}$. 
Using the metric form (\ref{Exp:ADMmetric}), 
we can express the action as  
\begin{eqnarray}
 S &=
\frac{\Mp^2}{2} \int \sqrt{h} \Bigl[ N\,\sR - 2 N
  V(\phi) + N (\kappa_{ij} \kappa^{ij} - \kappa^2) \cr
  & \qquad \qquad \qquad \qquad + \frac{1}{N} ( \dot{\phi}
  - N^i \partial_i \phi )^2 - N h^{ij} \partial_i \phi \partial_j \phi
  \Bigr] \dd^4x~, \label{Eq:ADMaction}
\end{eqnarray}
where $\sR$ is the three-dimensional Ricci scalar, and $\kappa_{ij}$ and
$\kappa$ are the extrinsic curvature and its trace, defined by 
\begin{eqnarray}
 \kappa_{ij} = \frac{1}{2 N} ( \dot{h}_{ij} - D_i N_j
 - D_j N_i )~, \qquad  
 \kappa = h^{ij} \kappa_{ij} ~.
\end{eqnarray}
The spatial indices $i, j, \cdots$ are raised or lowered by the spatial
metric $h_{ij}$ and $D_i$ denotes the covariant differentiation 
associated with $h_{ij}$. Since $N$ and $N^i$ are 
the Lagrange multipliers,  varying the action
with respect to them yields 
\begin{eqnarray}
 && \sR - 2 V -  (\kappa^{ij} \kappa_{ij} - \kappa^2 ) 
          -  N^{-2} ( \dot{\phi} - N^i \partial_i \phi)^2 - h^{ij}
 \partial_i \phi \partial_j \phi = 0~,   \\
 && D_j ( \kappa^j_{~i} - \delta^j_{~i} \kappa ) - N^{-1} 
  \partial_i \phi~ ( \dot{\phi} - N^j \partial_j \phi) = 0~,
\end{eqnarray}
which are called the Hamiltonian and 
momentum constraint equations, respectively. As we will show below 
more explicitly,
we eliminate the Lagrange
multipliers $N$ and $N_i$ from the action, solving these constraint equations.

As for the gauge conditions, we fix the time slicing 
by adopting the uniform field gauge:
\begin{eqnarray}
 \delta \phi =0\,.     \label{GC}
\end{eqnarray}
To impose spatial gauge conditions, we decompose 
the spatial metric $h_{ij}$ as
\begin{eqnarray}
 & h_{ij} = e^{2(\rho+\zeta)} \left[ e^{\delta \gamma} \right]_{ij}\,,  \label{Exp:metric}
\end{eqnarray}
where $a\equiv e^\rho(t)$ is the background scale factor, 
$\zeta$ is the so-called curvature perturbation and 
$\delta \gamma_{ij}$ is traceless, {\it i.e.}, 
$\delta {\gamma^i}_i=0$.
As the spatial gauge condition, we impose
the transverse conditions on $\delta \gamma_{ij}$:
\begin{eqnarray}
 & \partial_i \delta \gamma^i\!_j=0\, . 
\label{TTgauge}
\end{eqnarray}
We defer discussing the contribution from the tensor perturbation 
to Sec.~\ref{Sec:GW}.

Next, we eliminate the Lagrange multipliers $N$ and $N_i$ to derive the
action written in terms of the dynamical field $\zeta$ alone. 
In the above choice of the gauge, the constraint equations are given by
\begin{eqnarray}
 && \sR - 2 V -  (\kappa^{ij} \kappa_{ij} - \kappa^2 ) 
 - N^{-2}  \dot{\phi}^2 = 0\,,   \\ 
 && D_j  ( \kappa^j_{~i} - \delta^j_{~i} \kappa )  = 0 ~.
\end{eqnarray}
By expanding the metric perturbations as 
\begin{eqnarray}
  & \zeta= \zeta_I + \zeta_2 + \dots\,, \cr
  & N= 1 + N_1 + N_2 + \cdots\,,  \\
  & N_i = N_{i, 1} + N_{i, 2} + \cdots\,, \nonumber 
\end{eqnarray}
where $\zeta_I$ is the interaction picture field of $\zeta$ and 
the subscripts indicate the order of perturbation,   
the $n$-th order Hamiltonian and 
momentum constraints are expressed in the form
\begin{eqnarray}
 & V N_n - 3 \dot{\rho} \dot{\zeta}_n + e^{-2\rho}
 \partial^2 \zeta_n + \dot{\rho} e^{-2\rho} \partial^i N_{i,n} =H_n\,, 
\label{HCn}\\
& 4 \partial_i \left( \dot{\rho} N_n - \dot{\zeta}_n \right) -
 e^{-2\rho} \partial^2 N_{i, n} + e^{-2\rho} \partial_i
 \partial^j N_{j, n} = M_{i, n}\,,  \label{MCn}
\end{eqnarray}
where $\partial^2$ denotes the spatial Laplacian. $H_1$ and
$M_{i,1}$ are $0$ and $H_n$ and $M_{i,n}$ with $n\geq 2$ 
are functions which consist of $n$ interaction picture fields $\zeta_I$. 
To obtain the $n$-th order action,
we need to solve the constraint equations up to $[n/2]$-th order, 
where the square brackets mean the Gauss's floor function.

Since the constraint equations (\ref{HCn}) and
(\ref{MCn}) are elliptic-type, we need to
employ (spatial) boundary conditions to fix a solution 
for $N_n$ and $N_{i, n}$. 
As was shown in Appendix of Ref.~\cite{SRV2}, 
Eqs.~(\ref{HCn}) and (\ref{MCn}) give 
\begin{eqnarray}
  N_n & = \frac{1}{\dot{\rho}} \dot{\zeta}_n + \frac{V}{4 \dot{\rho}}
   e^{-2\rho} \left( e^{2\rho} \partial^{-2} \partial^i M_{i, n} -   G_n
	      \right) \,,
\label{Exp:Nn} \\
  N_{i, n} &= \partial_i \partial^{-2} \biggl[
 \frac{\dot{\phi}^2}{2\dot{\rho}^2} e^{2\rho}  \dot{\zeta}_n -
 \frac{1}{\dot{\rho}} \partial^2 \zeta_n 
+ \frac{e^{2\rho}}{\dot{\rho}} H_n - \frac{V}{4\dot{\rho}^2}
  \left\{ e^{2\rho}
   \partial^{-2} \partial^j   M_{j, n}  - G_n \right\} 
  \biggr] \cr
 & \qquad  - \left( \delta_i\!^j - \partial_i \partial^{-2}
 \partial^j  \right) \left\{ e^{2\rho} \partial^{-2}  \left( M_{j, n} -
		      \frac{4\dot{\rho}}{V} \partial_j H_n \right)  - G_{j, n}
 \right\}\,, \label{Exp:Nin}
\end{eqnarray}
where the degrees of freedom for the boundary conditions 
are manifestly expressed by adding homogeneous 
solutions $G_n(x)$ and $G_{i, n}(x)$ that satisfy
\begin{eqnarray}
 \partial^2 G_n(x) = 0\,, \qquad \quad \partial^2 G_{i, n}(x) =0\,. 
\end{eqnarray}
Since the function $G_{i, n}(x)$ contributes
only through its transverse part, we see that the number of introduced
independent functions is three. By employing appropriate boundary conditions at the spatial infinity,
the degrees of freedom of the boundary conditions for these elliptic-type equations will be uniquely
fixed. Substituting thus obtained expressions for $N$ and $N_i$, 
the action $S=\int \dd^4 x {\cal L}[\zeta,\, N,\, N_i]$ 
can be, in principle, expressed only in terms of the dynamical 
field $\zeta$. Then, the
evolution of $\zeta$ is governed by a {\it non-local} action which
contains the inverse Laplacian.

\subsection{The non-interacting theory and the scale-invariant spectrum} \label{SSec:free}
In this subsection, we consider the linear theory for the curvature
perturbation, which describes the evolution of 
the interaction picture field $\zeta_I$. 
For brevity, we introduce the horizon flow
functions as  
\begin{eqnarray}
 & \varepsilon_1\equiv - \frac{1}{\dot{\rho}} \frac{\dd}{\dd \rho} \dot{\rho}\, ,\qquad \quad
 \varepsilon_{n}\equiv \frac{1}{\varepsilon_{n-1}} \frac{\dd}{\dd \rho}
 \varepsilon_{n-1}~,
\end{eqnarray}
with $n \geq 2$, but we do not assume these functions are small.   
The quadratic action is given by
\begin{eqnarray}
 &  S_0 = \Mp^2 \int  \dd t\, \dd^3 \bm{x} \,
e^{3\rho} \varepsilon_1
 \biggl[ \dot\zeta_I^2  -
 e^{-2\rho} ( \partial_i \zeta_I)^2 \biggr] \,,  \label{Exp:S0}
\end{eqnarray} 
and the equation of motion for $\zeta_I$ is given by
\begin{eqnarray}
 & \left[  \partial^2_t + \left( 3 + \varepsilon_2 \right) \dot{\rho}
    \,\partial_t - e^{-2\rho}\partial^2 \right] \zeta_I(x)=0 \,. \label{Eq:GR}
\end{eqnarray}

For quantization, we expand $\zeta_I(x)$ as 
\begin{equation}
  \zeta_I(x)= \int \frac{\dd^3 \bm{k}}{(2\pi)^{3/2}}  a_{\sbm{k}} v_k(t)
   e^{i \sbm{k}\cdot \sbm{x}} + ({\rm h.c.})  \,, \label{Exp:expsi}
\end{equation}
with the mode function $v_k$ that satisfies 
\begin{eqnarray}
 \left[  \partial^2_t + \left( 3 + \varepsilon_2 \right) \dot{\rho}
    \,\partial_t + e^{-2\rho} k^2 \right] v_k(t) = 0\,.  \label{Eq:vk}
\end{eqnarray}
The mode function is normalized as  
\begin{eqnarray}
 & \left( v_k e^{i \sbm{k} \cdot \sbm{x}},\, v_p e^{i \sbm{p} \cdot
 \sbm{x}} \right) = (2\pi)^3 \delta^{(3)} (\bm{k} - \bm{p})\,, \label{Cond:N}
\end{eqnarray}
where the Klein-Gordon inner product is defined by 
\begin{eqnarray}
  (f_1, f_2) \equiv 
 - 2 i \Mp^2 \int \dd^3 \bm{x}\,  e^{3\rho} \varepsilon_1 
  \{ f_1 \partial_t f_2^* - (\partial_t f_1) f_2^*\}\,.
\end{eqnarray}
With this normalization, the commutation relations for $\zeta_I$
and its conjugate momentum yield those for the
creation and annihilation operators as
\begin{eqnarray}
 & \left[ a_{\sbm{k}},\, a_{\sbm{p}}^\dagger \right] =
 \delta^{(3)}(\bm{k}- \bm{p}), \qquad \left[ a_{\sbm{k}},\,
 a_{\sbm{p}} \right] = 0\,.
\end{eqnarray}
Using Eq.~(\ref{Exp:expsi}), we obtain the Wightman function of
$\zeta_I$ for the vacuum defined by $a_{\sbm{k}} |0 \rangle =0$ as
\begin{eqnarray}
 & G^+(x_1,\, x_2)= \langle 0| \zeta_I(x_1) \zeta_I(x_2) |0 \rangle 
 = \int \frac{\dd^3 \bm{k}}{(2\pi)^3} e^{i\sbm{k} \cdot
 (\sbm{x}_1-\sbm{x}_2)}  v_k(t_1) v_k^*(t_2) \,. 
\end{eqnarray}

A couple of comments are in order regarding the mode function
$v_k(t)$. As an implementation of the three-dimensional 
general covariance, the curvature 
perturbation is ensured to be massless, which
can be also confirmed in Eq.~(\ref{Eq:vk}). Because of that, when the
physical wavelength $ \lambda_{\rm phys} \sim e^\rho/k$ becomes much
longer than the Hubble scale, {\it i.e.}, 
$\lambda_{\rm phys} \dot{\rho} \sim  e^\rho \dot{\rho}/k \gg 1$, the
growing mode of the mode equation (\ref{Eq:vk}) rapidly approaches a
constant as
\begin{eqnarray}
 \frac{1}{\dot{\rho}} \partial_t v_k(t) 
= {\cal O} \left( (k/e^\rho\dot{\rho})^2 \right) v_k(t)\,.  \label{vkIR}
\end{eqnarray}
Choosing a solution of the mode equation selects the
vacuum state for the system with the interaction turned off. 
In the inflationary universe, 
the physical wavelength should be much shorter than the 
Hubble scale in the distant past. 
In this limit the mode function approximately behaves as that of a 
harmonic oscillator with a constant frequency with respect to the 
conformal time, 
\begin{equation}
\eta(t) \equiv \int^t \frac{\dd t'}{e^{\rho(t')}}~. \label{Def:eta}
\end{equation} 
Then, we can solve Eq.~(\ref{Eq:vk}) using the WKB approximation 
with the asymptotic boundary condition 
\begin{eqnarray}
  v_k(t) \to  \frac{1}{\Mp \sqrt{2 \varepsilon_1} e^{\rho(t)}} \frac{1}{
   \sqrt{2k}} e^{- i k \eta(t)}\,, \qquad \quad {\rm for}\quad - k \eta(t)
   \to \infty\,. \label{vkUV}
\end{eqnarray}
The vacuum state defined by this WKB solution is called the adiabatic
vacuum. 
When the background spacetime is approximated by the de Sitter space,
which is the case for $k/(e^\rho \dot{\rho}) \agt 1$ in the slow roll 
inflation, 
the mode function for the adiabatic vacuum is reduced to 
\begin{eqnarray}
  v_k(t) \approx   \frac{i}{
   \sqrt{2k^3}} \frac{1}{\sqrt{2 \varepsilon_1}} \left( \dot{\rho} \over
						  \Mp \right) \{ 1 + i k
   \eta(t) \} e^{- i k \eta(t)}\,. \label{vkBD}
\end{eqnarray}
In this case, 
the power spectrum becomes 
almost scale-invariant in the IR limit as 
\begin{eqnarray}
  P(k) \equiv |v_k(t)|^2 = \frac{1}{4 k^3} \frac{1}{ \varepsilon_1(t_k)}
   \left( \dot{\rho}(t_k) \over \Mp \right)^2  \left[ 1 + {\cal O} \left(
  (k \eta)^2 \right) \right]\,,  \label{Exp:Pk}
\end{eqnarray}
where we evaluated $v_k(t)$ at the Hubble crossing time $t=t_k$ with
$k=e^{\rho(t_k)} \dot{\rho}(t_k)$, since 
the curvature perturbation gets frozen rapidly after the time $t_k$.

\subsection{Various types of divergences} \label{SSec:Vdiv}
Now, we consider 
the $n$-point functions of the curvature perturbation $\zeta$, 
turning on the interaction. 
Using the in-in (or equivalently the closed time path) formalism~\cite{Calzetta:1986cq}, 
the $n$-point function for $\zeta$ is calculated as    
\begin{eqnarray}
 \left\langle  \zeta(t, \bm{x}_1) \cdots \zeta(t, \bm{x}_n)
 \right\rangle  = \left\langle
 U_I^\dagger(t,\, t_i)  \zeta_I(t, \bm{x}_1)  \cdots  \zeta_I(t, \bm{x}_n)  
 U_I(t,\, t_i) \right\rangle, \label{Exp:np} 
\end{eqnarray}
where $t_i$ is an initial time and 
\begin{eqnarray}
 & U_I(t_1,\, t_2) = T \exp \left[- i\int^{t_1}_{t_2} \dd t\,\int \dd^3
			     \bm{x} {\cal H}_I(t,\, \bm{x})
			    \right]\,,  
\end{eqnarray}
is the unitary operator with the interaction Hamiltonian density 
${\cal H}_I(x)$, which consists of the interaction
picture field $\zeta_I$. Using Eq.~(\ref{Exp:np}), we can expand the
$n$-point functions for $\zeta$ in terms of the Wightman propagator
$G^{\pm}(x_1,\, x_2)$. 

A naive computation of the $n$-point functions 
tells us that loop integrals of 
perturbations in an inflationary spacetime apparently 
have various kinds of unsuppressed 
contribution from the deep IR modes. 
In this subsection, we illustrate and classify potential
origins of such pathological behaviors. 
The first three (discussed in Sec.~\ref{IRtypeI} - \ref{IRtypeIII}) 
are related to the momentum integrals, while 
the last (discussed in Sec.~\ref{IRtypeIV}) is originating from the time integral. 
Here, our illustration is focusing on
the curvature perturbation in single field models, 
but almost the same arguments will 
follow also for the tensor perturbation.

\subsubsection{The IR divergence}
\label{IRtypeI}
\begin{figure}
\begin{center}
\includegraphics[width=10cm]{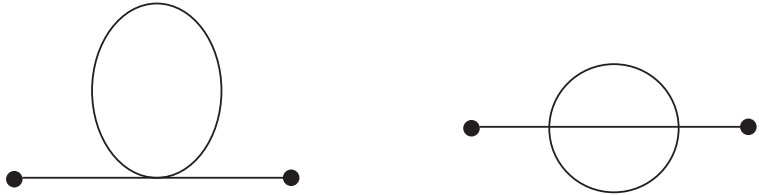}
\caption{Diagrams which potentially yield the IRdiv.}  
\label{Fg:loops}
\end{center}
\end{figure}

When we assume that the corresponding free theory has an almost scale invariant spectrum in
the IR limit, a naive consideration can easily lead to the IR divergence
due to loop corrections. To explain this, we pick
up the following quartic interaction vertex:
\begin{eqnarray}
  {\cal H}_I(x) \ni  \{\zeta_I(x)\}^2 \left\{  \frac{\partial_i}{e^{\rho}
   \dot{\rho}} \zeta_I(x) \right\}^2 
\label{Exp:exIR}
\end{eqnarray}
from the interaction Hamiltonian density ${\cal H}_I(t)$, where we abbreviated
 unimportant time-dependent coefficients. 
Using the in-in formalism, we find that 
the one-loop diagram depicted in Fig.~\ref{Fg:loops}
obtained from the contraction between the two $\zeta_I$s in 
the interaction vertex~(\ref{Exp:exIR}) yields the factor
\begin{eqnarray}
 G^+(x,\, x) = \langle \{ \zeta_I(x)\}^2 \rangle = \int \frac{\dd^3
  \bm{k}}{(2\pi)^3}  |v_k(t)|^2.  
\end{eqnarray}
We can easily understand that 
this momentum integral logarithmically diverges 
in the IR as $\int \dd^3 \bm{k}/k^3$ for the scale invariant spectrum. 
Even if the spectrum is not completely scale invariant as given in 
Eq.~(\ref{Exp:Pk}), deep IR modes 
contribute to $\langle \zeta^2_I \rangle$ significantly. 
We refer to the appearance of such an unsuppressed momentum 
integral for small $k$ as {\em IR divergence} (IRdiv), even though the 
integral does not diverge for the blue spectrum. 
Note that we encounter the same IRdiv also in a free theory, when we
evaluate the spectrum in the position space. 

\subsubsection{The IR secular growth}
\label{IRtypeII}
\begin{figure}
\begin{center}
\includegraphics[width=10cm]{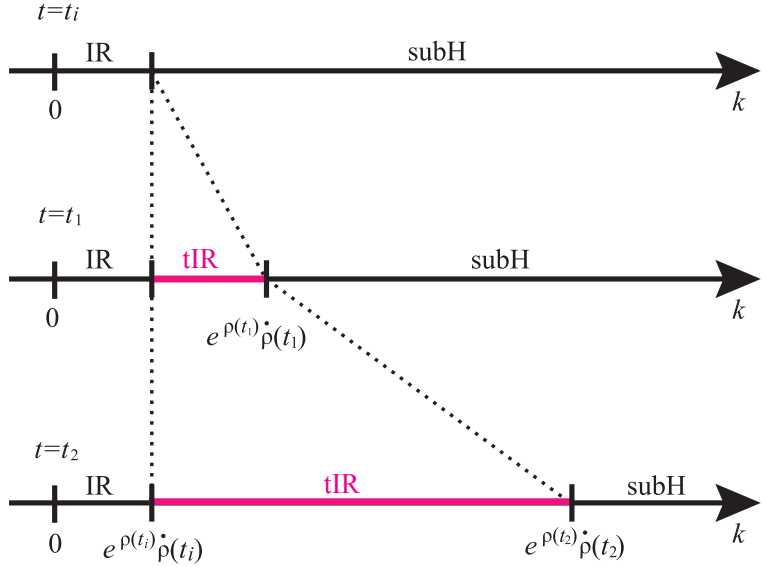}
\caption{Classification of the wavenumbers $k$ into the IR modes, tIR
 modes, and subH modes.}  
\label{Fg:tIR}
\end{center}
\end{figure}

One may think of regularizing the IRdiv by introducing an IR
cutoff. When we introduce the IR cutoff, say at the Hubble scale for 
the initial time $t_i$, the variance of the super Hubble (superH) modes:
\begin{eqnarray}
  \langle \{\zeta_I(x)\}^2 \rangle_{\rm superH} \propto \int^{e^{\rho(t)}
   \dot{\rho}(t)}_{e^{\rho(t_i)} \dot{\rho}(t_i)} 
   \frac{\dd k}{k} = \ln \left\{\frac{e^{\rho(t)}
			  \dot{\rho}(t)}{e^{\rho(t_i)} \dot{\rho}(t_i)} \right\}  \label{Eq:secular}
\end{eqnarray}
shows the secular growth which is logarithmic in the scale factor
$a=e^\rho$. Then, the loop corrections, which are suppressed by an extra
power of the amplitude of the power spectrum  
$(\dot{\rho}/\Mp)^2$, may dominate in case inflation continues
sufficiently long, leading to the breakdown of perturbation. We refer to 
the modes with
$e^{\rho(t_i)} \dot{\rho}(t_i) \alt k \alt e^{\rho(t)} \dot{\rho}(t)$
as {\it the transient IR} (tIR) modes and refer to the enhancement of the loop
contributions due to the tIR modes as {\it the IR
secular growth} (IRsec), discriminating it from IRdiv. To be precise, we define 
the tIR modes as such that were in the sub Hubble (subH) range at the initial
time $t_i$, 
but were transmitted into superH ones by the time $t$ 
as shown in Fig.~\ref{Fg:tIR}. 
As inflation proceeds, the range of the tIR modes increases, which leads to 
the IRsec. Equation (\ref{Eq:secular}) shows that the introduction of an 
artificial comoving IR cutoff eliminates IRdiv, but it does not cure IRsec.

Note that the IRsec manifestly depends on the value of the IR cutoff. 
When we denote the comoving wavenumber for the IR cutoff as $k_{\IR}$, 
the secular growth factor is given by 
$\ln (e^\rho \dot{\rho}/k_{\IR})$. In Ref.~\cite{Lyth2007}, Lyth
discussed the cutoff dependence of the correlation functions and
proposed to set the IR cutoff length scale to a slightly larger scale than the observable universe, {\it i.e.}, 
$k_{\IR} \sim e^{\rho_0} \dot{\rho}_0$ where the subscript $0$ 
indicates the quantities evaluated at the the present epoch
(see also Ref.~\cite{Lyth2006}). This cutoff dependence was 
studied to the two-loop order by Bartolo
{\it et al.} in Refs.~\cite{Bartolo:2007ti, Dimastrogiovanni:2008af}. More recently, introducing an IR
cutoff $k_{\rm IR}$, Byrnes {\it et al}. examined the influence of the IRsec on the non-Gaussian parameters
in Refs.~\cite{Tasinato:2012js, Nurmi:2013xv, Byrnes:2013qjy}.
Although introducing an IR cutoff might give an 
almost correct practical prescription, it should be verified to be a
proper way to compute observable quantities. As Enqvist {\it et al.} pointed out, the present Hubble
scale is not a critical scale of the theory beyond 
which the presence of fluctuations is prohibited 
and hence the introduction of
the IR cutoff at $k_{\IR} \sim e^{\rho_0}\dot{\rho}_0$ 
is ad hoc~\cite{Enqvist:2008kt}.

The potential IRsec has been also addressed from the point 
of the gradual change of effective coupling constants. 
A particular interest is in the screening of the
cosmological constant. As an example of the secular change 
of an effective coupling constant, we consider how the cosmological 
constant is affected by the loop correction due to 
a massless scalar field with a quartic potential 
$V(\phi) = \lambda \phi^4/4!$ in a
fixed quasi de Sitter background. 
By picking up only the tIR modes, one may naively
expect that the variance of
the free massless field would be given by 
$\langle \phi^2 \rangle_{{\rm tIR}} \simeq (\dot{\rho}/\Mp)^2 \ln
\{e^{\rho(t)} \dot{\rho}(t)/e^{\rho(t_i)} \dot{\rho}(t_i)\}$.
If we could simply trust this expression, the expectation value of the 
potential term in the energy-momentum tensor 
would be evaluated as
\begin{eqnarray}
 \langle T_{\mu \nu} \rangle \ni - g_{\mu \nu} \langle V(\phi)
  \rangle \simeq - \lambda g_{\mu \nu}  \left\{  \left(\dot{\rho} \over
						  \Mp \right)^2 \ln
  \left ( e^{\rho(t)} \dot{\rho}(t) \over e^{\rho(t_i)} \dot{\rho}(t_i) \right)  \right\}^2~,
\label{antiscreen}
\end{eqnarray}
signaling the time-dependence of the cosmological constant. 
In this simple example, the cosmological constant increases but 
the screening may happen when we consider different 
field contents~\cite{TW96, TW962,ABM}. 
(For the detail of computation, see Ref.~\cite{Woodard:2005cw}
and references therein.)

Of course, the secular change of coupling 
constants due to the superH modes should be examined more carefully. 
The evolution of the superH modes can be naively understood based on
the stochastic approach, which was initiated by
Starobinsky~\cite{S}, while the quantum loop effect is not essential
there. As we shall discuss in more detail in 
Sec.~\ref{SSec:Mass}, 
in this approach the evolution of the field value 
averaged over the Hubble scale, 
$\bar\phi$, is 
described as a stochastic motion caused by the successive 
addition of modes transmitted 
from the subH modes to the tIR modes. 
This stochastic diffusion balances with the deterministic force 
pushing the average value toward the bottom of the potential in the end. 
As a result, the variance of the massless scalar field with the quartic
potential approaches 
$\langle \bar\phi^2 \rangle \to \dot{\rho}^2/(\sqrt{\lambda} \Mp^2)$ after a
sufficiently long time~\cite{SY}. 
If we start with $\bar \phi=0$, the stochastic diffusion dominates 
and the $\bar\phi$ deviates from the bottom of the potential. 
One can understand that 
$\langle \phi^2\rangle_{\rm tIR}$ increases in time because of this effect.
Then, the energy momentum tensor in each horizon patch will 
naturally have the value corresponding to $V(\bar\phi)$,  
whose ensemble average will give the result in Eq.~(\ref{antiscreen}).
However, the local physics in each Hubble patch is still described by
the original $\lambda\phi^4$ model with the stochastic background value
of $\bar{\phi}$. This stochastic interpretation mentioned above suggests that the secular change 
of coupling constants obtained by explicit calculations 
does not necessarily mean that the accumulated IR modes can 
modify the local physics law.

In the above discussion 
the metric perturbations have been neglected. Once
we include them, we also need to pay attention to the gauge issue, 
which is our main focus of this review paper. 

\subsubsection{The inverse Laplacian}
\label{IRtypeIII}
Another complication may arise from the inverse Laplacian operators, 
$\partial^{-2}$,  
contained in the expression for the lapse function $N$ and 
the shift vector $N_i$, (\ref{Exp:Nn}) and (\ref{Exp:Nin}). 
Using these expressions for $N$ and $N_i$, we see that the interaction Lagrangian 
written in terms of the curvature perturbation $\zeta$ 
contains $\partial^{-2}$. These inverse Laplacian operators are always 
associated with at least two derivative operators in the action,
because $\partial^{-2}$ in $N$ explicitly accompanies
two spatial derivative operators, while $\partial^{-2}$ in $N_i$
accompanies at least one and $N_i$ is always multiplied by at least one
spatial derivative operator in the action.
For example, besides the unimportant time-dependence, 
the interaction Hamiltonian contains a term like 
\begin{equation}
  {\cal H}_I(x) \ni {1\over \dot \rho^2}\dot\zeta_I(x) \partial^i \zeta_I(x) 
    \partial^{-2}\left(\dot\zeta_I(x) \partial_i\zeta_I(x) \right). \label{tempint}
\end{equation}
Using this vertex, we can consider the one-loop diagram 
as shown in the left panel of Fig.~\ref{Fg:loops}, which 
yields the factor in Fourier components
\begin{eqnarray}
 \int \dd^3 \bm{k}_1  {\bfk \cdot\bfk_1 \over |\bfk_1-\bfk_1|^{2}}
      \partial_t G^+_{\sbm{k}_1}(t,t') \big|_{t' \to t}~.      
\end{eqnarray}
This factor is very pathological. For any value of $\bfk_1$ the 
integrand is divergent. A different kind of pathology may appear from the two-loop diagram 
shown in the right panel of Fig.~\ref{Fg:loops}. If we use the
interaction vertex in Eq.~(\ref{tempint}) twice, the diagram yields the factor
in Fourier components 
\begin{eqnarray}
 \int \dd^3 \bm{k}_1 \dd^3 \bm{k}_2 {(\bfk_1\cdot\bfk_2)^2\over |\bfk-\bfk_1|^{4}}
     G^+_{\sbm{k}_1}(t,t') G^+_{\sbm{k}_2}(t,t') 
    \partial_t\partial_{t'}G^+_{\sbm{k}-\sbm{k}_1-\sbm{k}_2}(t,t')~. 
\end{eqnarray}
Unless some non-trivial cancellation occurs, the contribution 
at around $\bfk_1=\bfk$ of this diagram diverges. We need to make sure
that the inverse Laplacian operator does not give rise to a singular
pole in the momentum integral.

\subsubsection{The secular growth due to temporal integral}
\label{IRtypeIV}
The remaining issue is the possible secular growth (SG) due to the
accumulated contribution from the temporal integral. If the contribution to some observable quantity 
from the interaction vertex in the far past
remains unsuppressed, it will diverge when we 
send the initial time to the infinite past. 
We think that the discrimination of this effect from 
the previously introduced IRsec is important. 
The main difference is in that the IRsec can 
be discussed without taking into account subH modes, 
specified by $k \agt e^{\rho(t)} \dot{\rho}(t)$, while 
the SG in general can be caused by 
the contribution from vertex composed of subH modes. 
In Refs.~\cite{Weinberg05, Weinberg06}, Weinberg investigated the SG from
the time integration, performing the time integral with the momenta of
the propagators fixed. He assumed that
the mode function in the limit $k \gg e^{\rho(t)} \dot{\rho}(t)$
oscillates very rapidly and hence the subH modes 
$k \agt e^{\rho(t)} \dot{\rho}(t)$ give only little contribution. This
assumption will not be verified for an arbitrary initial
state. Actually, in general, a time integration includes a mixture of the positive and
negative frequency mode functions, which yields 
the phase in the UV limit $e^{i \eta(t) (k_1 - k_2 + k_3 - \cdots)}$. 
Then, the phase does not necessarily exhibit the rapid
oscillation even for the modes with $ - k_m \eta(t) \gg 1$. In Sec.~\ref{Sec:Euclidean}, we
will show that when we fix an initial state by employing the so
called $i\epsilon$ prescription, the assumption of 
rapid oscillations is satisfied.

The subH modes also include ultraviolet (UV) modes with $k \gg e^{\rho(t)} \dot{\rho}(t)$. 
We refer to the divergence due to the UV modes as {\it the UV divergence}. 
In Refs.~\cite{Weinberg05, SZ09}, the UV divergence has been identified
by using the dimensional regularization. Initially it was pointed out that 
the integral over the subH modes can also contribute to the SG, but this SG is shown to be 
an artifact by means of a consistent dimensional regularization~\cite{SZ09}. 
In this review article, therefore, we will not provide a rigorous
argument about the UV regularization. 

\subsection{The IR divergences in QED and QCD}  \label{SSec:QEDQCD}
It is widely known that the IR divergence also appears in
QED or non-Abelian gauge theories. A frequently asked 
question is whether the IR divergences in these gauge theories in the flat spacetime 
has something to do with the IR pathologies discussed in the previous
subsection. 

In the case of the IR divergence in gauge theories, 
we can discretize the singular pole by
using the dimensional regularization, because changing the spacetime
dimension from $D=4$ to $D=4 - \delta$ with a negative $\delta$ reduces
the power in the IR, relaxing the singular behaviours in the IR limit. 
By contrast, in the cosmological setup, changing the spacetime
dimensions does not change the behavior in the IR. For example, in the $D$-dimensional de Sitter space, the 
power spectrum of a massless scalar field is given by 
$P(k) \propto 1/k^{D-1}$, and hence the logarithmic divergence
remains as~\cite{Janssen:2008px} 
$$
\int \frac{\dd^{D-1}\bm{k}}{(2\pi)^{D-1}} P(k) \propto 
 \int \frac{\dd k}{k}\,,
$$
indicating that the dimensional regularization cannot
regularize the logarithmic divergence associated with the IR
contributions. 

In QED and QCD, the IR divergences from the vertex corrections are
cancelled in the cross section by the ones from the soft photon and
gluon radiations, respectively (see Ref.~\cite{QCD} and references
therein). This is an implementation of the Kinoshita-Lee-Nauenberg
theorem, which states that in a theory with massless fields, the soft
divergence should be cancelled in transition rates, if we sum over the
initial and final degenerate states. Here the degeneracy means that 
an electron accompanied by an arbitrary number of soft photons 
cannot be distinguished from a single electron in an experiment.
The roll played by the soft photons might be attributed to the 
IR modes in the perturbation of the inflationary universe. 
It is analogous to the property of soft photons 
that the IR modes are hardly detected by the local observers. 
In fact, Seery suggested an analogy between
the Fokker-Planck equation in the stochastic approach~\cite{S} and the
equation which describes the evolution of the parton distribution functions in
Ref.~\cite{Seery:2009hs}. Hence, a similarity between these two cases
might be worthy of examination.

However, 
the origin of divergence in QED 
is the neglection of the states with soft photons. 
By contrast, in the calculation of cosmological 
perturbation, what we calculate as observable quantities are the
correlators in a particular quantum state. In this case the degeneracy
due to the IR modes in the final state is not neglected, because
all possible final states are automatically summed up in the in-in
formalism. On the other hand, as for the initial state, it might be still suggestive
to claim by analogy that all the IR fluctuations must be added to the
initial state in a proper way to avoid the IRdiv and IRsec, but
the precise meaning of this speculative statement is not so clear.

\subsection{The dilatation symmetry}   \label{SSec:Dilatation}
The regularization of the IRdiv and IRsec, which are both caused by 
superH modes, has been attempted, based on various 
methods~\cite{IRgauge_L, IRgauge,  SRV1, SRV2, IRsingle, IRmulti,  BGHNT10, GHT11, GS10, GS11,
GS1109,SZ1203, PSZ, SZ1210,  IRgauge_multi}. In this subsection, we give a brief 
review on the previous works, focusing on 
the dilatation symmetry that 
must be fully taken into account in proving the absence of the IRdiv
and IRsec. 

As is expected from the fact that the
spatial metric is given in the form $e^{2(\rho+\zeta)}  \dd \bm{x}^2$,
a constant shift of the dynamical variable $\zeta$ can be absorbed by 
the overall rescaling of the spatial coordinates. 
Hence, the action for $\zeta$ preserves the dilatation symmetry:
\begin{eqnarray}
 & x^i \to e^{-s} x^i\,, \qquad  \zeta(t,\, \bm{x}) \to
 \zeta(t,\, e^{-s} \bm{x}) - s\,,  \label{Exp:dilatation}
\end{eqnarray}
where $s$ is a constant parameter.  (In 
literature this dilatation symmetry has been addressed many times. 
See, for instance, Refs.~\cite{Creminelli:2012ed, Hinterbichler:2012nm} and the
references therein.) One may naively expect that
we can remove the divergent IR contribution in $\zeta$ 
using this constant shift. 
In fact, if we set the parameter $s$ to $\bar{\zeta}(t_i)$, the 
averaged value of $\zeta$ over the Hubble patch at $t_i$, the
logarithmically divergent Wightman function would be regularized. For instance, its coincidence limit, 
$\langle \{\zeta_I(t,\,\bm{x})\}^2 \rangle$, 
would be replaced with $\langle \{\zeta_I(t,\,\bm{x}) -
\bar{\zeta}_I(t_i)\}^2 \rangle$, whose superH modes give 
\begin{eqnarray}
 \langle \{\zeta_I(t,\, \bm{x}) - \bar{\zeta}_I(t_i)\}^2  \rangle_{\rm superH} 
\propto  \int^{e^{\rho(t)}
   \dot{\rho}(t)}_{e^{\rho(t_i)} \dot{\rho}(t_i)} 
   \frac{\dd k}{k} \,,
\end{eqnarray}
where the comoving radius of the Hubble patch 
is given by $1/(e^{\rho(t_i)} \dot{\rho}(t_i))$. As we discussed at around Eq.~(\ref{Eq:secular}), 
although the introduction of the comoving IR cutoff eliminates the
IRdiv, it does not eliminate the IRsec. 

One may think that if the system can be described in such a way that the
symmetry under the time-dependent dilatation transformation is manifest, 
the logarithmic growth of $\langle \{\zeta_I(t,\,\bm{x}) -
\bar{\zeta}_I(t_i)\}^2 \rangle$ might be eliminated by setting $s(t)$ to the time-dependent spatial
average in the Hubble patch. 
However, the reduced action written in terms of $\zeta$
does not preserve the invariance under the dilatation 
transformation with $s(t)$ being time dependent.
For example, in Ref.~\cite{Hinterbichler:2012nm}, the
authors showed that when we consider the whole universe with the
infinite spatial volume, the dilatation transformation should be 
time independent to preserve the action invariant. 
In addition, the $n$-point
functions with the inverse Laplacian $\partial^{-2}$ in the 
interaction vertexes do not seem to be
regularized merely by considering the dilatation symmetry. This quick
consideration indicates that
the presence of the dilatation symmetry 
may play an important role to show the absence of the IRdiv and IRsec, 
but it is not enough to resolve these pathologies.

\section{The causality and the gauge invariance}   \label{Sec:GI}
Our goal is to judge whether or not the various divergences mentioned in
the preceding section are just artifacts. 
In this paper, we will show that the actual observable quantities are
not spoiled by these divergences. In this section, we will provide the
basic ingredients in the discussion. We will also clarify that a short proof 
based on naive arguments is quite unsatisfactory. 

\subsection{Influence from the causally disconnected region}   \label{SSec:Influence}
\begin{figure}
\begin{center}
\includegraphics[width=10cm]{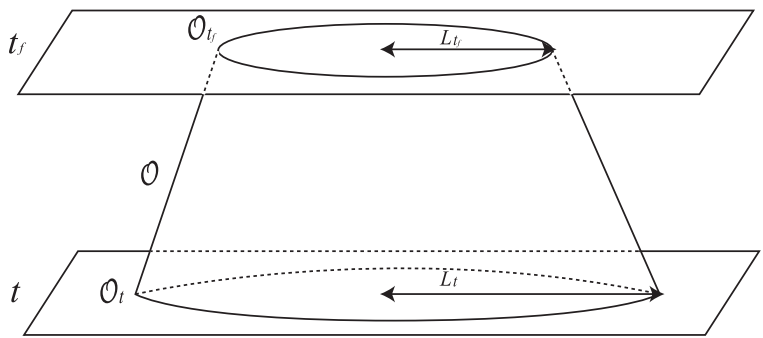}
\caption{The observable region.}  
\label{Fg:OR}
\end{center}
\end{figure}

First, we define the observable region
as the region causally connected to us. We denote the observable region 
on the time slicing at the end of inflation $t_f$ and 
its comoving radius as ${\cal O}_{t_f}$ and $L_{t_f}$, respectively. 
The causality
requires that $L_{t_f}$ should satisfy
$L_{t_f} \alt \int^{t_0}_{t_f} \dd t/e^{\rho(t)}$,
where $t_0$ is the present time. 
What we will detect through the
cosmological observations will be the $n$-point functions 
of the fluctuation with the arguments $(t_f, \, \bm{x})$ 
contained in the observable region
${\cal O}_{t_f}$. For later use, we refer to 
the causal past of ${\cal O}_f$ as the
observable region ${\cal O}$ and refer to the intersection between 
${\cal O}$ and a $t$-constant slicing $\Sigma_t$ as
${\cal O}_t$ (see Fig.~\ref{Fg:OR}). 
We approximate the comoving radius of the region ${\cal O}_t$ as
\begin{eqnarray}
   L_t \equiv L_{t_f} + \int^{t_f}_t \frac{\dd t'}{e^{\rho(t')}} 
 \simeq L_{t_f} + \frac{1}{e^{\rho(t)} \dot{\rho}(t)}\,,
\end{eqnarray}
which approaches the comoving Hubble radius, 
$1/e^{\rho(t)} \dot{\rho}(t)$, in the distant past. 

One can argue that the effects of the superH modes 
with $k \alt e^{\rho(t)} \dot{\rho}(t)$ 
are the influence from the outside of the
observable region ${\cal O}$. 
These modes potentially affect the fluctuations in ${\cal O}_{t_f}$ by two ways. 
One is due to the non-local interaction through 
the inverse Laplacian $\partial^{-2}$, while 
the other is through the Wightman function $G^+(x_1,\, x_2)$. 
Even if the spatial distance 
$|\bm{x}_1- \bm{x}_2|$ is bounded from above by confining $\bm{x}_1$ and
$\bm{x}_2$ within the observable region, the contribution 
to $G^+(x_1,\, x_2)$ from IR modes with 
$k \leq |\bm{x}_1- \bm{x}_2|^{-1}$ are not suppressed. 
These modes make $G^+(x_1,\, x_2)$ divergent for scale-invariant or red-tilted 
spectrum. To regularize the contribution from the superH modes, we
need to prove the suppression of their effects.

\subsection{The residual coordinates degrees of freedom in the local
  universe}  \label{SSec:RDF}
In the previous subsection, we introduced the observable region 
${\cal O}$, which is a limited portion of the whole universe. We claim
that the observable fluctuation must be composed of fluctuations in
${\cal O}$. Furthermore, since the information that we can access 
is limited to within ${\cal O}$, 
there is no reason to request the regularity at the
spatial infinity in solving the elliptic constraint equations
(\ref{HCn}) and (\ref{MCn}), at least, at the level of Heisenberg 
equations of motion. 
Then, there arise degrees of freedom in choosing the boundary
conditions, which appear as arbitrary 
homogeneous solutions of the Laplace equation, 
$G_n(x)$ and $G_{i,n}(x)$ in Eqs.~(\ref{Exp:Nn}) and (\ref{Exp:Nin}). 
These arbitrary functions in $N$ and $N_i$ can be understood as 
the degrees of freedom in choosing the spatial coordinates. 
Since the time slicing 
is fixed by the gauge condition (\ref{GC}), the residual gauge degrees
of freedom only reside in the spatial coordinates $x^i$.

As we have shown in
Refs.~\cite{IRgauge_L, IRgauge}, these residual coordinate
transformations associated with $G_n(x)$ and $G_{i, n}(x)$ 
are expressed as
\begin{eqnarray}
 &  x^i \to x^i - \sum_{m=1}^\infty {s^i}_{j_1 \cdots j_m}(t) x^{j_1}
 \cdots x^{j_m} + \cdots\,, \label{Exp:RC} 
\end{eqnarray}
where ${s^i}_{j_1 \cdots j_m}(t)$ are symmetric traceless tensors, which
satisfy 
$\delta^{j j'} {s^i}_{j_1 \cdots j \cdots j'  \cdots j_m}(t)=0$. 
Here, we abbreviated the non-linear terms in Eq.~(\ref{Exp:RC}). 
These transformations diverge at the spatial
infinity, no matter how small the coefficients are. 
By contrast, restricted to the local
region, the magnitude of the coordinate transformations~(\ref{Exp:RC})
is kept perturbatively small. 
Since the transformations~(\ref{Exp:RC}) 
are nothing but 
coordinate transformations, the Heisenberg equations of motion for the
diffeomorphism invariant theory remains unchanged 
under these transformations. 
Note that these coordinate transformations include the
dilatation transformation with the time dependent function $s(t)$.

We should note that, once we substitute the expressions for $N$ and $N_i$
to obtain the equation of motion solely written in terms of 
the curvature perturbation $\zeta$, 
the symmetry under the residual coordinate transformations is
lost, because $N$ and $N_i$ depend on the 
specified boundary conditions. 
Although in this sense the coordinate transformations~(\ref{Exp:RC}) 
are to be distinguished from the usual gauge transformation 
that leaves the overall action invariant, 
we are accustomed to call infinitesimal coordinate transformations 
gauge transformation. To avoid confusion, we distinguish 
the coordinate transformations~(\ref{Exp:RC}) as the \gauge
transformation by using the italic font.  

\subsection{The IR issues and changing the local average}  \label{SSec:LA}
In the previous subsection, we pointed out the presence of 
residual \gauge degrees of freedom 
from the point of view restricted to the local observable universe. 
Among them, here we focus
on the dilatation transformation $x^i \to e^{-s(t)} x^i$, denoting the
trace part of $s^i\!_j(t)$ as $s(t)$. 
As we mentioned earlier, 
this dilatation transformation shifts 
the spatial curvature perturbation as 
$\zeta(t,\, \bm{x}) \to \zeta(t,\, e^{-s(t)} \bm{x}) - s(t)$, and 
hence it can be understood as subtraction of the local
average in the observable region ${\cal O}_t$ 
from $\zeta$. (As shown in Refs.~\cite{IRgauge_L, IRgauge}, one of the residual \gauge
transformations (\ref{Exp:RC}) can absorb the local average of the
tensor perturbation as well.) The \gauge invariance will imply that 
the quantities that we can observe in actual measurements should
be insensitive to this change of the local average of $\zeta$. 

As far as we know, computing the local observable quantity was first
emphasized in the discussion of the long
wavelength fluctuations by Unruh~\cite{Unruh}. Afterwards,
Geshnizjani and Brandenberger examined the behaviour of the
long wavelength fluctuations by considering a local quantity~\cite{GB02}. 
They computed the local expansion rate $\Theta \equiv {u^\mu}_{;\mu}$
where $u^\mu$ denotes the time-like four vector which is orthogonal to
the $\phi$-constant hypersurfaces, {\it i.e.},
$u_\mu = \partial_\mu \phi /\sqrt{\partial^{\nu} \phi \partial_\nu \phi}$
 in a single clock inflation. They showed
that when the local expansion rate $\Theta$ is evaluated as a function 
of the clock field $\phi$, $\Theta$ is not affected by 
long wavelength modes, basically staying at the background value, 
$\Theta \simeq \sqrt{V(\phi) /3 }$. By contrast, 
$\Theta$ as a function of the cosmological time $t$ suffers from 
the logarithmic secular growth discussed in
Sec.~\ref{SSec:Vdiv}. Their analysis is totally classical, but their
result suggests that the accumulation of the superH modes may
disappear, if we evaluate genuinely \gauge invariant quantities. 
(See also their discussion in two field models~\cite{GB03}.)

In Refs.~\cite{BGHNT10, GHT11, GS10}, the leading IR logarithms of the
curvature perturbation are discussed focusing on the dilatation transformation, which introduces
the shift of $\zeta$. In these references, the time-independent
dilatation is addressed, but here we extend it to the time-dependent one to consider both the IRdiv and IRsec. In Refs.~\cite{BGHNT10, GHT11}, the authors
introduced the spatial average of the curvature perturbation in the
Hubble patch with the size $L_t \sim 1/(e^\rho \dot{\rho})$,  
which is roughly expressed as
$
\bar{\zeta}(t) \sim \int_{|k|< e^\rho \dot{\rho}} \dd^3 \bm{k}\, \zeta_{\sbm{k}}(t) 
$
 in terms of the Fourier components $\zeta_{\sbm{k}}$. 
As inflation proceeds, the
number of the modes that contribute to $\bar{\zeta}(t)$ 
increases, 
leading to the secular growth of $\bar{\zeta}(t)$. 
Because of the contribution of $\bar{\zeta}(t)$, the
physical meaning of the comoving coordinates $\bm{x}$ 
is effectively modified as $e^{\bar{\zeta}(t)} \bm{x}$. 
The IRdiv and IRsec computed in the $\delta N$ formalism are shown to agree with the
divergent contributions which appear from the above modification of the
physical distance due to $\bar{\zeta}(t)$. 
In Ref.~\cite{GS10}, a similar argument
is provided based on a semi-classical approach.

\subsection{How to fix the residual \gauge degrees of freedom}
In the previous subsection, we claimed that the IRdiv and IRsec are
deeply related to the presence of the residual \gauge degrees
of freedom. In this subsection, we discuss several attempts to show the absence of the IRdiv and 
IRsec by fixing the residual \gauge degrees of freedom. The discussions in this
subsection will not complete a rigorous proof of the IR regularity, 
but giving an overview of these attempts will be instructive 
to capture a key aspect that should be taken into account 
in proving the absence of the IRdiv and IRsec. 

\subsubsection{Absorbing the IR divergence by gauge fixing}   \label{SSSec:GF}
One way to preserve the invariance under the gauge transformation is
fixing the gauge conditions completely. The residual
\gauge degrees of freedom explained above can be also 
removed by employing additional \gauge conditions 
that fix the boundary
conditions for $N$ and $N_i$ at the boundary of the local 
region ${\cal O}$. 
Then, the IR regularity may be explicitly shown by performing the quantization in this
local region, since the wavelengths that fit within this local
region ${\cal O}$ are bounded by the size of the region ${\cal O}$. 
Although the quantization in the local region might be an interesting idea, 
it is not clear how to select a natural initial quantum state for the system 
after the removal of the residual \gauge degrees of freedom. 
Even the spatial translation symmetry of 
the quantum state cannot be easily guaranteed in this approach, 
because it is manifestly broken by the boundary conditions imposed at a finite distance.

In Ref.~\cite{IRsingle}, to preserve the global translation symmetry 
in the spatial directions manifestly, 
the initial state is set at an initial time $t=t_i$ 
without fixing the residual \gauge degrees of freedom. 
Then, a shift of
the spatial coordinates $\bm{x}$ to $\bm{x}+\bm{a}$ is simply 
absorbed by multiplying the overall phase factor $e^{i\sbm{k} \cdot \sbm{a}}$ 
to each Fourier mode. 
After setting the initial quantum state in this way, 
the residual \gauge transformation (\ref{Exp:RC}) is performed
to absorb the IR contributions. 
In this approach one can show the absence of the IRdiv and IRsec,  
if they are absent at the initial time, which however 
is not guaranteed. Below we briefly summarize the discussion in Ref.~\cite{IRsingle}.

As usual, in quantizing the curvature perturbation $\zeta(x)$, 
we first consider the whole universe.  
The initial conditions for $\zeta(x)$ and the conjugate momentum $\pi(x)$ are set by 
\begin{eqnarray}
\zeta(t_i, \bm{x}) \equiv  \zeta_I(t_i, \bm{x})\,, \qquad 
\pi(t_i, \bm{x}) \equiv  \pi_I(t_i, \bm{x}) \,,  \label{Exp:ICzeta}
\end{eqnarray}
with the corresponding interaction picture fields $\zeta_I(x)$ and $\pi_I(x)$.  
The mode expansion of $\zeta_I(x)$ is given in Eq.~(\ref{Exp:expsi}). 
Then, we perform the dilatation transformation $x^i \to e^{-s(t)}x^i$,
which is one of the residual \gauge transformation, and 
the curvature perturbation transforms as
\begin{eqnarray} 
 & \zeta(t,\, \bm{x}) \to \zeta'(t,\, \bm{x})= \zeta(t,\,
 e^{-s(t)} \bm{x}) - s(t) \,. \label{Exp:RC2}
\end{eqnarray}
We fixed the time dependent parameter $s(t)$, requesting
\begin{eqnarray}
  \bar{\zeta}'(t) =0\,, \label{Exp:LGC}
\end{eqnarray}
where $\bar{\zeta}'(t)$ is the local spatial average in
the observable region, defined by
\begin{eqnarray}
 \bar{\zeta}'(t) \equiv \frac{\int \dd^3 \bm{x} W_t (\bm{x}) \zeta'(t,\,
  \bm{x})}{\int \dd^3 \bm{x} W_t (\bm{x})} \,,
\end{eqnarray}
with a window function $W_t(\bm{x})$ which is 
non-vanishing only in the local region ${\cal O}_t$. Then $\zeta'(x)$
becomes the 
curvature perturbation in the gauge more 
relevant to the local observable universe.
Here we recast the discussion 
given in the flat gauge in Ref.~\cite{IRsingle} 
into the one in the uniform field gauge.

The issue of the inverse Laplacian addressed in Sec.~\ref{IRtypeIII},
can also be solved by using the remaining residual
\gauge degrees of freedom. 
If we choose the boundary conditions for 
$\partial^{-2}$ in $N$ and $N_i$ appropriately, 
$N$ and $N_i$ in the region ${\cal O}_t$
can be specified by the fluctuations only within ${\cal O}_t$. 
In the general solutions of $N$ and $N_i$ given in
Eqs.~(\ref{Exp:Nn}) and (\ref{Exp:Nin}), the residual \gauge
degrees of freedom are expressed by arbitrary homogeneous 
solutions of the Laplace equation, $G_n(x)$ and 
$(\delta_i\!^j - \partial_i \partial^{-2} \partial^j)G_{j, n}(x)$. 
We fix the homogeneous solution $G_n(x)$, requesting that
$\partial^{-2}\partial^i M_{i, n}(x)$ in $N_n$ should satisfy
\begin{eqnarray}
 &  - \frac{1}{4 \pi} \int \frac{\dd^3 \bm{y}}{|\bm{x} - \bm{y}|}
    W_t(\bm{y}) \partial^i M_{i, n}(t,\, \bm{y}) =  
\partial^{-2}\partial^i M_{i, n}(x) 
-   e^{-2\rho} G_n (x) \,, 
\label{Exp:LBC}
\end{eqnarray}
in the observable region ${\cal O}_t$.
Similarly, using the transverse part of $G_{i, n}(x)$, we can fix the
boundary conditions for the remaining $\partial^{-2}$ 
so as to eliminate the influence from the region 
far outside of ${\cal O}_t$. 
(For a detailed explanation, see Appendix of Ref.~\cite{SRV2}.) 
Then, all the interaction vertexes are 
confined to the neighborhood of ${\cal O}$. 

After the gauge fixing, 
the Heisenberg equation for $\zeta'(x)$, 
which is perturbatively expanded as 
$\zeta'(x)=\zeta'_I(x) + \zeta'_2(x) + \cdots$, 
can be iteratively solved as
\begin{eqnarray}
 \zeta'_n(t,\, \bm{x}) = \breve{\zeta}_n(t,\, \bm{x}) - \frac{\int \dd^3
  \bm{x} W_t(\bm{x}) \breve{\zeta}_n(t,\, \bm{x})}{\int \dd^3 \bm{x}
  W_t(\bm{x})}   \label{Exp:zetafixed}
\end{eqnarray}  
where $\breve{\zeta}_n(x)$ is given by
\begin{eqnarray}
  \breve{\zeta}_n (x) = - 2 \Mp^2 \int \dd t' \int \dd \bm{x}'
   \varepsilon_1(t') e^{3\rho(t')} G_R(x,\, x') \Gamma_n(x')\,, \label{Exp:zetafixedbreve}
\end{eqnarray}
with the retarded Green function that satisfies 
\begin{eqnarray}
 & \left[  \partial^2_t + \left( 3 + \varepsilon_2 \right) \dot{\rho}
    \partial_t - e^{-2\rho}\partial^2 \right] G_R(x,\, x')=  -
 \frac{1}{2 \Mp^2} \frac{1}{\varepsilon_1 e^{3\rho}} \delta^{(4)}(x - x')\,. \label{Eq:GR}
\end{eqnarray}
Here, the interaction picture fields $\zeta_I'(x)$ and its conjugate
momentum $\pi_I'(x)$ satisfy the initial conditions
\begin{eqnarray}
 \zeta'(t_i,\, \bm{x}) \equiv \zeta_I'(t_i, \bm{x})\,, \qquad
 \pi'(t_i,\, \bm{x}) \equiv \pi_I'(t_i, \bm{x})\,, 
\label{Def:zetad}
\end{eqnarray}
and $\Gamma_n(x)$ 
denotes the non-linear interaction terms that 
include $n$ $\zeta_I'(x)$s. After we fixed the boundary condition of the
inverse Laplacian as above, $\Gamma_n(x)$ becomes a local function which
is not affected by the far outside of ${\cal O}_t$. Reflecting the fact
that the retarded Green function $G_R(x,\, x')$ takes a non-vanishing
value only if the two points $x$ and $x'$ are causally connected, the
above expression manifestly preserves the (approximate) causality.  
Note that the local average of $\zeta'_n(x)$, given by operating
$\int \dd^3 \bm{x}\, W_t(\bm{x})$ on Eq.~(\ref{Exp:zetafixed}), vanishes
as is requested from the gauge condition (\ref{Exp:LGC}).

Using Eqs.~(\ref{Exp:zetafixed}) and (\ref{Exp:zetafixedbreve}), we can
expand the correlation functions for the curvature perturbation
$\zeta'(x)$ in terms of the retarded Green function and the 
correlation functions for $\zeta'_I$. 
Since the integration region of the vertex
integrals are restricted to the local observable region 
${\cal O}$, if the correlation functions for $\zeta'_I$ were all finite, 
those for $\zeta'(x)$ would be finite as well. Here, however, we should
notice that $\zeta'_I$ is not linear in the original interaction picture
field $\zeta_I(x)$.

The gauge condition (\ref{Exp:LGC}) implies that 
the part of $\zeta'(x)$ which is linear in $\zeta_I(x)$ 
should be given by 
\begin{eqnarray}
  \zeta_L'(t,\, \bm{x}) =  \zeta_I(t,\, \bm{x}) -\bar{\zeta}_I(t) \,, \label{Exp:zetaLd}
\end{eqnarray}
where $\bar{\zeta}_I(t)$ denotes the local average of $\zeta_I$, given by
\begin{eqnarray}
  \bar{\zeta}_I(t) \equiv  \frac{\int \dd^3
  \bm{x} W_t(\bm{x}) \zeta_I(t,\, \bm{x})}{\int \dd^3 \bm{x}
  W_t(\bm{x})}\,.
\end{eqnarray}
Inserting the mode expansion of $\zeta_I(x)$, given in
Eq.~(\ref{Exp:expsi}) into Eq.~(\ref{Exp:zetaLd}), 
the two-point function for $\zeta_L'(x)$ is calculated as 
\begin{eqnarray}
\hspace*{-2cm}
 \langle \zeta_L'(x_1) \zeta_L'(x_2) \rangle 
 = \int \frac{\dd^3 \bm{k}}{(2\pi)^3} 
\left( e^{i \sbm{k} \cdot \sbm{x}_1} -
 \frac{\hat{W}_t(- \bm{k})}{\hat{W}_t(0)} \right)
 \left( e^{- i \sbm{k} \cdot \sbm{x}_2} -
 \frac{\hat{W}_t(\bm{k})}{\hat{W}_t(0)} \right) v_k(t_1) v^*_k(t_2)   \label{Exp:zetad2p}
\end{eqnarray}
where $\hat{W}_t(\bm{k})$ denotes the Fourier mode of the window
function $W_t(\bm{x})$. Since 
the contributions of the IR and tIR modes to $\zeta_I(x)$ 
are canceled by the second term in
Eq.~(\ref{Exp:zetaLd}), 
the IR suppression factor
\begin{eqnarray}
   \left| e^{i \sbm{k} \cdot \sbm{x}} -
 \frac{\hat{W}_t(- \bm{k})}{\hat{W}_t(0)} \right| \alt {\cal O} \left( k
 L_t \right)\,,  \label{tempsup}
\end{eqnarray}
arises. 
As a result, the momentum integration in the two-point function
for $\zeta'_L(x)$ is regularized in the IR. 
To obtain the inequality (\ref{tempsup}), we used the facts 
that the spatial coordinates $\bm{x}$ satisfy
$|\bm{x}| \alt L_t$ and that $\hat{W}_t(\bm{k})$ can be
expanded as $\hat{W}_t(\bm{k})/\hat{W}_t(0) = 1 + {\cal O}(k L_t)$.

However, the regularity of $\langle \zeta_L'(x_1) \zeta_L'(x_2) \rangle$
does not imply the regularity of the correlation functions 
for $\zeta'_I(x)$, because $\zeta'_I(x)$ also contains non-linear terms
of $\zeta_I(x)$. Inserting 
Eqs.~(\ref{Exp:ICzeta}) and (\ref{Def:zetad}) into Eq.~(\ref{Exp:RC2}),
we find that $\zeta'_I(x)$ 
and $\zeta'_L(x)$ are related as
\begin{eqnarray}
 &\zeta_I'(t_i,\, \bm{x}) = \zeta'_L(t_i,\, \bm{x}) - \bar{\zeta}_I(t_i) \{  \bm{x} \cdot \partial_{\sbm{x}}
  \zeta_I(t_i,\, \bm{x}) - \overline{\bm{x} \cdot \partial_{\sbm{x}} \zeta_I} (t_i)
  \} +  {\cal O}(\zeta_I^3)\,,  \label{Exp:ID}
\end{eqnarray}
where $\overline{\bm{x} \cdot \partial_{\sbm{x}} \zeta_I}(t)$ denotes the local
average of $\bm{x} \cdot \partial_{\sbm{x}} \zeta_I(x)$. Since the non-linear
terms in Eq.~(\ref{Exp:ID}) include $\bar{\zeta}_I(t_i)$ which diverges due
to the IR modes, they can make the correlation functions 
for $\zeta'_I(x)$ divergent. The lesson is that in general it is not
straightforward to absorb all the IR
divergent contributions by performing the residual \gauge transformation.

We should also note that, if 
we can eliminate the non-linear terms in Eq.~(\ref{Exp:ID}), 
$\zeta'_I$ 
agrees with $\zeta'_L$ 
at the initial time. 
Both variables satisfy the same equation of motion, 
{\it i.e.}, the linearized equation of motion 
with the subtraction of the local average as given in 
Eq.~(\ref{Exp:zetafixed}).  
Therefore, in this case the correlation functions of $\zeta'_I$ 
can be replaced by the products of 
the two-point function of $\zeta'_L$,  
$\langle \zeta_L'(x_1) \zeta_L'(x_2) \rangle$, which is 
shown above to be regular in IR. 
Hence, the regularity of the correlation functions 
for $\zeta'(x)$ follows. We will see that the IR regularity
conditions, which will be derived in Sec.~\ref{SSec:IRdivC}, 
agree with the conditions
which request the non-linear terms in Eq.~(\ref{Exp:ID}) should vanish.

In Ref.~\cite{SZ1203}, to absorb the IR modes, Senatore and
Zaldarriaga introduced a physical distance measure, considering the map
between a scale in comoving coordinates $L$ and the Hubble crossing time
of the corresponding scale $t_{hc}$, which satisfies  
$\dot{\rho}(t_{hc})^{-3} = e^{3\{\rho(t_{hc}) + \bar\zeta(t_{hc})\}} L^3$,
where $\bar{\zeta}(t_{hc})$ denotes the spatial average of the curvature
perturbation in the region with the size $L$. They computed the
physical volume at the reheating $t_{rh}$, eliminating the influence of
the IR modes which resides in the comoving coordinates as
$$
V_{rh} = e^{3\rho(t_{rh})} \int \dd^3 \bm{x} e^{3\zeta(t_{rh},\,
\sbm{x})} = e^{3\{\rho(t_{rh})-\rho(t_{hc})\}} \dot{\rho}(t_{hc})^{-3} 
  \int \frac{\dd^3 \bm{x}}{L^3} e^{3\{\zeta(t_{rh},\,\sbm{x})- \bar{\zeta}(t_{hc})\}}\,.
$$
If we could replace all $\zeta$s in $V_{rh}$ with the interaction picture field
$\zeta_I$, we would see that the IR modes $k L \ll 1$ in 
$\zeta_I(t_{rh},\,\bm{x})$ is canceled by these modes in
$\bar{\zeta}_I(t_{hc})$. In a non-linear computation, the
discussion will become more complicated because choosing the proper
measure does not necessarily guarantee that all the interaction picture
fields $\zeta_I$ appear in the combination $\zeta_I - \bar{\zeta}_I$,
but their argument still suggests that the choice of the proper measure is one of
the crucial ingredients for the regularization of the IR
contributions. A related aspect was focused also in
Refs.~\cite{Enqvist:2008kt, Page}. In Ref.~\cite{Enqvist:2008kt},
focusing on the fact that the IRdiv is canceled in the difference
between the free propagators 
$
\langle \zeta_I(t,\, \bm{x}_1) \zeta_I(t,\, \bm{x}_2) \rangle - 
 \langle \zeta_I(t,\, \bm{x}_1) \zeta_I(t,\, \bm{x}_3) \rangle\,,
$
Enqvist {\it et al.} computed a non-linear quantity which partially
includes the one-loop contributions but whose IRdiv is canceled.
We also comment on the paper by Primentel {\it et al.}~\cite{PSZ}. They discussed
fixing the residual \gauge degrees of freedom in a different
way from the one described above. 
The argument was extended to higher order loops in Ref.~\cite{SZ1210}. 
They performed the residual \gauge transformation
$x^i \to M^i\!_j(t) x^j + C^i(t)$, requesting that 
$N_i$ and $\partial_jN_i$ 
after the \gauge transformation 
should vanish in the local region at the
leading order in $kL$, where $\bm{k}$ is the wavenumber assigned to these
variables. If this requirement can be fulfilled, the Fourier 
modes of these variables $N_{i, \sbm{k}}$ and $k_j N_{i, \sbm{k}}$ will
obtain an additional suppression in the IR by $kL$. In Refs.~\cite{PSZ,
SZ1210}, they gave a restricted analysis on the \gauge fixing, 
picking up the term with $\partial_i \partial^{-2} \dot{\zeta}_{n}$ in 
$N_{i,n}$ as an example.

\subsubsection{Sending the initial time to past infinity} 
The appearance of the IRdiv due to the residual \gauge transformation mentioned in Sec.~\ref{SSSec:GF}
might be evaded by sending the initial time $t_i$ to the past infinity. 
This is because in this limit 
the size of the observable region ${\cal O}_t$ in
comoving coordinates, $L_t$, becomes infinitely large. 
As we get close to this limit, 
the discrepancy between the average in the local region and 
that in the whole universe becomes smaller and smaller. 
Then, the residual \gauge transformation at the initial
time might be unnecessary.  
Also, it is quite natural to eliminate the presence of 
a special time in setting the initial quantum state. 

We should note that when we send the initial time to the past infinity, 
it is too naive to 
neglect the subH 
modes with $k \agt e^{\rho(t)} \dot{\rho}(t)$ at the distant
past. The discussion in Sec.~\ref{SSSec:GF} suggests that if 
the non-linear correlations are regular at a finite reference time, they
will be kept regular also for the later times. However, the non-linear correlations at a finite reference time 
cannot be specified without knowing the evolution 
in the subH regime up to that time. This aspect makes the IR issues very complicated.

To get an intuition, let's consider the conservation of $\zeta_{\sbm{k}}$ in the limit 
$k/(e^\rho \dot{\rho}) \ll 1$ as an example. This conservation is a well
known fact in the long-wavelength approximation. (For a liner analysis,
see Ref.~\cite{WMLL} and for a non-linear extension, see Ref.~\cite{LMS} and references therein.)   
However, once we include the non-linear contributions from 
the subH modes, the conservation does not hold any more. If we consider
a vertex integral confined to the region ${\cal O}$ as in
Sec.~\ref{SSSec:GF}, the superH modes will be suppressed but the subH
modes can still contribute. Since the domain of the time
integration is infinite, it is easy to understand that there is a possible
origin of the SG due to the subH modes, and the effects may diverge in the 
limit $t_i \to - \infty$.

In Refs.~\cite{SZ1210, ABG}, the absence of the SG was
claimed relying on the conservation of the curvature perturbation, but the
aspect mentioned above has not been discussed. 
In addition, even if the conservation of 
$\zeta_{\sbm{k}}$ in the limit $k/(e^\rho \dot{\rho}) \ll 1$ is proved, 
the logarithmic enhancement in the form 
$(k/e^\rho \dot{\rho})^2 \ln \{k/ e^{\rho(t_i)} \dot{\rho}(t_i)\}$ may
give rise. The factor $\ln \{k/ e^{\rho(t_i)} \dot{\rho}(t_i)\}$ can
become large to overcome the suppression by $(k/e^\rho \dot{\rho})$ 
when we send the initial time to the past infinity. 

\subsection{Constructing a gauge invariant operator}
The observable fluctuations should be free from 
the residual \gauge degrees of freedom, 
which were introduced in Sec.~\ref{SSec:RDF}. 
In this subsection, following Refs.~\cite{IRgauge_L, IRgauge}, we 
construct an operator which is invariant under the residual \gauge
transformations. We call such an operator a genuinely \gauge invariant operator.  

\subsubsection{The definition}      \label{SSSec:Def}
Since the time slicing is uniquely specified 
by the gauge condition (\ref{GC}), a quantity which is
invariant under the transformation of spatial coordinates will be
genuinely \gauge invariant. 
To construct a genuinely \gauge
invariant operator, we propose to calculate
$n$-point functions for the scalar curvature of the induced metric on 
a $\phi\!=$constant hypersurface, $\sR$. 
Although $\sR$ itself transforms as a scalar quantity, 
the $n$-point functions of $\sR$ 
with its $n$ arguments specified in a
coordinate-independent manner will be gauge invariant. 
The distances of spatial geodesics 
that connect pairs of $n$ points characterize the configuration 
in a coordinate independent manner. 
Based on this idea, we specify the $n$ spatial points in terms of the 
geodesic distances and the directional cosines, measured from 
a reference point. 
Although the reference point and frame depend on the coordinates, 
this would not matter as long as we choose
a quantum state that respects the spatial 
homogeneity and isotropy of the universe.

The geodesic normal coordinates on each time slice
are introduced by solving the spatial geodesic equation 
\begin{eqnarray}
 \frac{\dd^2 x_{gl}^i}{\dd \lambda^2} +  {^s \Gamma^i}_{jk} \frac{\dd
  x_{gl}^j}{\dd \lambda} \frac{\dd x_{gl}^k}{\dd \lambda} =0~,
\label{GE}
\end{eqnarray}
where ${^s \Gamma^i}_{jk}$ is the Christoffel symbol with respect to 
the three dimensional spatial metric and
$\lambda$ is the affine parameter. 
The affine parameter ranges
from $\lambda=0$ to $1$, and the initial ``velocity'' is given by
\begin{eqnarray}
 \frac{\dd x^i_{gl}(\bm{x},\lambda)}{\dd \lambda} \bigg\vert_{\lambda=0}= e^{-\zeta(\lambda=0)}
 x^i\,. \label{IC}
\end{eqnarray}
Here we associate the subscript $gl$ with the global coordinates, 
reserving the simple notation $\bm{x}$ for the geodesic normal 
coordinates. 
A point $\bm{x}$ in the geodesic normal coordinates is mapped 
to the end point of the geodesic $x_{gl}^i(\bm{x},\lambda=1)$ 
in the original coordinates. 
We perturbatively expand $x_{gl}^i$ in terms of $x^i$ as 
$$
x_{gl}^i= x^i + \delta x^i(\bm{x}).
$$ 
With the aid of the geodesic normal coordinates, 
we can construct 
a genuinely \gauge invariant variable as 
\begin{eqnarray}
 {^g\!R}(x) &\equiv \sR (t,\, x_{gl}^i(\bm{x})) = 
 \sR (t,\, x^i + \delta x^i (\bm{x})))\,. \label{Def:gR}
\end{eqnarray}

Since the gauge invariant variable $\gR$ 
does not include the conjugate momentum of $\zeta$, we can consider 
products of $\gR$ at an equal time without any ambiguity 
of operator ordering. To calculate the $n$-point functions of $\gR$,
we need to specify the quantum state as well. 
One may think that the quantum state should be selected so that it
preserves the invariance under the residual \gauge transformations. 
However, we cannot directly discuss this invariance as 
a condition for allowed quantum states in this approach, because the residual
\gauge degrees of freedom are absent when we quantize fields in the
whole universe. 

Here we note that even though the operator $\gR$ is not affected by the residual
\gauge degrees of freedom, this does not imply that 
the $n$-point functions of $\gR$ are uncorrelated to the fields 
in the causally disconnected region. In Sec.~\ref{SSec:Influence}, we
discussed two ways in which 
the variables in the observable region ${\cal O}$ 
are contaminated by the influence from the outside of 
${\cal O}$. 
Since changing the boundary condition for the inverse
Laplacian $\partial^{-2}$ can be thought of as a residual \gauge
transformation, selecting $\partial^{-2}$ whose integration region is 
restricted to the region ${\cal O}$ as in Eq.~(\ref{Exp:LBC}) does not affect the
$n$-point functions of $\gR$. 
Therefore,
as long as we consider a genuinely \gauge invariant operator,
the inverse Laplacian $\partial^{-2}$ never gives the conjunction between the inside and the outside of ${\cal O}$. 
On the other hand, the long-range correlation through the Wightman
function can stay even in the genuinely \gauge invariant variables, 
providing a possible origin of the IRdiv and IRsec. 
In Sec.~\ref{Sec:SRV}, we will show that requesting the
absence of the IRdiv or IRsec, in fact, constrains the 
quantum state of the inflationary universe. This can be interpreted as
that the IR regularity of observable fluctuations can be 
achieved only when the quantum state is selected 
so that the long-range correlation is suppressed for 
genuinely \gauge invariant variables.

\subsubsection{The coarse grained distance} 
Tsamis and Woodard~\cite{Tsamis:1989yu} showed that using the geodesic
normal coordinates can introduce an additional origin of UV
divergence, which may not be
renormalized by local counter terms~\cite{Miao:2012xc}.
This is expected because specifying the spatial distance precisely 
in the presence of the gravitational perturbation requires taking
account of all the short wavelength modes of the gravitational
perturbation. In any realistic observations, what we observe are smeared 
fields with a finite resolution. However, it is not so trivial how to introduce a realistic 
smearing in a gauge invariant manner. 
Here, just to keep the UV contributions under control, we replace
the geodesic normal coordinates 
with approximate ones removing 
the UV contributions. 
Originally the geodesic normal coordinates are related to $x^i_{gl}$
as
\begin{eqnarray}
 & x^i_{gl} = e^{-\zeta(t,\, e^{-\zeta} \sbm{x})}  x^i + \cdots\,,
\label{gnc0}
\end{eqnarray}
where the ellipsis means 
the terms
suppressed in the IR limit, which vanish 
when $\zeta(x)$ is spatially homogeneous. 
We replace the above relation with 
\begin{eqnarray}
 & x^i_{gl} = e^{-\Gbz(t)}  x^i\,, 
\label{gnc}
\end{eqnarray}
where we introduced the smeared curvature perturbation
\begin{eqnarray}
 & \Gbz(t) \equiv \frac{\int \dd^3 \bm{x}\, W_t(\bm{x}) \zeta(t,
 e^{- \Gbz}\bm{x})
 }{\int \dd^3 \bm{x}\, W_t(\bm{x})}\,.
\label{Def:Gbz} 
\end{eqnarray} 
Although $\Gbz$ appears on
the right-hand side of Eq.~(\ref{Def:Gbz}), 
this expression defines $\Gbz$ 
iteratively at each order of the perturbation. 
We calculate the $n$-point functions of 
${\cal R}_x \gz(t, \bm{x})$, instead of $\gR$, with 
\begin{eqnarray}
 & \gz (t,\, \bm{x}) 
\equiv \zeta (t,\,   e^{-\Gbz(t)} \bm{x})\,.
\label{Exp:gz}
\end{eqnarray}
Here, ${\cal R}_x$ represents such operators, 
\begin{eqnarray}
\frac{\partial_t}{\dot{\rho}},\quad \frac{\partial_{\sbm{x}}}{e^{\rho(t)}
 \dot{\rho}(t)},\quad \biggl(1 - \frac{\int \dd^3 \bm{x}
 W_t(\bm{x})}{\int \dd^3 \bm{y} W_t(\bm{y})} \biggr), \quad 
\cdots\,,
\end{eqnarray}
that manifestly suppress the IR contributions by acting on the field 
$\gz (t,\, \bm{x})$.

${\cal R}_x\gz(t,\, \bm{x})$ is not invariant under general residual \gauge transformations but 
still is invariant under the dilatation transformation,
which can absorb the dominant IR contributions. 
In fact, since the genuinely \gauge invariant variable $\gR(x)$ 
should be invariant under the dilatation transformation, 
$\gz(x)$ appears only in the form of ${\cal R}_x \gz(x)$ 
when we express $\gR(x)$ in terms of $\gz(x)$.
As we can compute $\gR(x)$ from ${\cal R}_x\gz(x)$, the $n$-point functions of $\gR$ should be
regular if those of ${\cal R}_x\gz$ are regular. Therefore, we compute
the latter.

Note that the genuinely \gauge invariant quantity $\gR(x)$ should be composed of
${\cal R}_x \gz(x)$, but this does not imply that 
all the interaction field operators in $\gR(x)$ are 
associated with IR suppressing operators ${\cal R}_x$ as 
${\cal R}_x \zeta_I(x)$. 
Therefore, as we mentioned at
the end of Sec.~\ref{SSSec:Def}, the long-range
correlation in the Wightman function $G^+(x,\, x')$ is not always suppressed. In
Sec.~\ref{Sec:Euclidean}, we will show that 
only for restricted quantum states 
all the Wightman functions contained in the expression for the
$n$-point functions of $\gR$ 
are accompanied with the IR suppressing operators.

\section{Restricting initial states from the \gauge
 invariance and the regularity}  \label{Sec:SRV}
In the previous section, we revealed that the IRdiv and IRsec
originate from the influence from the outside of the observable region, 
and this influence is mediated by 
the residual \gauge degrees of freedom 
from the view point of the local observable universe. 
To remove the influence of the residual \gauge degrees of freedom,  
it is essentially important to focus on the correlation functions of 
genuinely \gauge invariant operators.  
However, evaluating genuinely \gauge invariant operators is not
sufficient. Another important aspect is to 
choose the quantum state which is not affected by the residual \gauge
degrees of freedom. In other words, 
we need to choose a quantum state whose correlation with the outside of
the observable region is suppressed. In this section, we will show that the conditions for the
absence of the IRdiv and IRsec yield a non-trivial restriction on 
the quantum state.
Then, we will show that this condition can be interpreted
as the condition for the invariance of the quantum state 
under the dilatation transformation. In this section,
for an illustrative purpose, we employ a simple assumption that
the interaction is turned on at a finite initial time $t_i$.

\subsection{Restricting initial states from the absence of IRdiv and IRsec}   \label{SSec:SRIC}
In this subsection, we compute the two-point function of 
${\cal R}_x \gz(x)$ up to the one-loop order, and derive the condition 
on the initial state for the absence of the IRdiv and IRsec. 
Assuming that the interaction is turned on at the initial time $t_i$, 
we set 
\begin{eqnarray}
 & \zeta(t_i,\, \bm{x}) = \zeta_I(t_i,\, \bm{x})\,,\qquad \quad
  \pi(t_i,\, \bm{x}) = \pi_I(t_i,\, \bm{x})\,, \label{IC}
\end{eqnarray}
where $\pi_I$ is the conjugate momentum of the interaction picture field
$\zeta_I$. Here, we compute the two-point function by solving the
Heisenberg equation of motion for the curvature perturbation operator 
$\zeta$. 
Using the retarded Green function $G_R(x,\, x')$,
we obtain the solution of $\zeta$ that
satisfies the initial condition (\ref{IC}) as
\begin{eqnarray} 
 & \zeta (x) = \zeta_I(x)  +  {\cal L}_R^{-1} {\cal S}_{\rm NL}(x)
\label{SolR}
\end{eqnarray}
with 
\begin{eqnarray}
 & {\cal L}_R^{-1} {\cal S}_{\rm NL}(t,\, \bm{x})  \equiv - 2 \Mp^2 \int
  \dd^4 x' \varepsilon_1(t') e^{3\rho(t')}  G_R(x, x') {\cal S}_{\rm NL}(x')\,,
 \label{Exp:SolR}
\end{eqnarray}
where the explicit form of the non-linear source term ${\cal S}_{\rm NL}(x)$ 
will be given later. Evaluating Eq.~(\ref{Exp:SolR}) iteratively, 
we can obtain an expression for the curvature perturbation. 
Inserting thus obtained solution
$\zeta$ into Eq.~(\ref{Def:Gbz}), we can perturbatively compute $\gz(x)$ as
\begin{eqnarray}
 \gz(x) = \zeta_I(x) + \gz_2(x) + \gz_3(x) + \cdots \,,  \label{Exp:gzp}
\end{eqnarray}
where $\gz_n(x)$ represents the term that consists of $n$ interaction 
picture fields $\zeta_I$. Expanding the interaction
picture field $\zeta_I$ as Eq.~(\ref{Exp:expsi}), the initial
vacuum state is defined by
\begin{eqnarray}
  a_{\sbm{k}} |0 \rangle =0 \,.  \label{Def:vacuum}
\end{eqnarray}
The $n$-point functions computed by taking the 
expectation values of products of thus obtained $\gz(x)$ 
can be formally shown to agree with those calculated 
in the in-in formalism (see, for instance, Appendix of
Ref.~\cite{SRV1}).

Using Eq.~(\ref{Exp:gzp}), the one-loop contributions to the two-point
function of ${\cal R}_x \gz(x)$ are given by 
\begin{eqnarray}
  &\langle {\cal R}_{x_1}\! \gz(x_1) {\cal R}_{x_2}\! \gz(x_2)
   \rangle_{1 {\rm loop}}  \cr
  &= \langle {\cal R}_{x_1}\! \gz_2(x_1) {\cal R}_{x_2}\! \gz_2(x_2)
  \rangle  + \langle {\cal R}_{x_1}\! \zeta_I(x_1) {\cal R}_{x_2}\! \gz_3(x_2)
  \rangle + \langle {\cal R}_{x_1}\! \gz_3(x_1) {\cal R}_{x_2}\! \zeta_I(x_2)
  \rangle\, \nonumber \\ \label{Exp:1loop}
\end{eqnarray}
After we choose the boundary conditions for $\partial^{-2}$ as given in
Eq.~(\ref{Exp:LBC}), the inverse Laplacian does not enhance the singular
behaviour of the superH modes, and hence the IRdiv and IRsec can appear
only from the variance  
$$
\langle \bar{\zeta}_I^2(t) \rangle 
\simeq \int_{k \leq 1/L_t} \frac{\dd^3 \bm{k}}{(2\pi)^3} P(k)\,.
$$
If $\gz_2$ includes $\bar\zeta_I$, the first
term on the right side in Eq.~(\ref{Exp:1loop}) can give
$\langle \bar\zeta_I^2 \rangle$. Similarly, 
if $\gz_3$ includes $\bar\zeta_I^2$, the
second and third terms can give $\langle \bar\zeta_I^2 \rangle$. 
To make our discussion compact and
transparent, here, we pick up only the potentially divergent
contributions, which yield $\langle \bar\zeta_I^2 \rangle$.  
We introduce the symbol ``$\Approx$'' to denote the 
approximate equality neglecting the terms which do not yield 
$\langle \bar\zeta_I^2 \rangle$ at the one-loop 
level~\cite{IRgauge_L, IRgauge}.   

We can easily derive the non-linear action which is relevant for yielding 
$\langle \bar{\zeta}_I^2 \rangle$ as 
\begin{eqnarray}
 &  S \Approx \Mp^2 \int  \dd t\, \dd^3 \bm{x} \,
e^{3(\rho+\zeta)} \varepsilon_1
 \biggl[ (\partial_t \zeta)^2  -
 e^{-2(\rho+\zeta)} ( \partial_i \zeta)^2
 \biggr] \,,
\end{eqnarray} 
where the terms with more than two fields with differentiations, 
which do not give $\langle \zeta_I^2 \rangle$, are abbreviated. 
This approximate expression for the action also preserves the dilatation 
symmetry~(\ref{Exp:dilatation})~\cite{SRV1}. 

The variation of the above action gives 
the equation of motion as
\begin{eqnarray}
  \left[  \partial^2_t + \left( 3 + \varepsilon_2 \right) \dot{\rho}
    \partial_t - e^{-2\rho}\partial^2 \right] \zeta(x) = {\cal S}_{\rm NL}(x)
\end{eqnarray}
with
\begin{eqnarray}
 {\cal S}_{\rm NL}(x) \Approx e^{-2\rho} (e^{-2\zeta} - 1) \partial^2
  \zeta(x)  -  \delta(t - t_i ) (e^{3\zeta} - 1) \partial_t \zeta(x) \,,  \label{SolRIR}
\end{eqnarray}
where the last term is added to satisfy the second condition in
Eqs.~(\ref{IC})~\cite{SRV1}. By inserting Eq.~(\ref{SolRIR}) into 
Eq.~(\ref{SolR}), the solution that satisfies Eq.~(\ref{IC}) 
is obtained. Then, we can express $\gz(x)$ as
\begin{eqnarray}
 & \gz_2(x) \Approx - \bar{\zeta}_I\, {\cal D}_x \zeta_I \,,  \qquad 
 \gz_3(x) \Approx \frac{1}{2} \bar{\zeta}_I^2\, {\cal D}_x^2 \zeta_I\,, 
\label{Exp:gz2}
\end{eqnarray}
with
\begin{eqnarray}
 {\cal D}_x \equiv   2 {\cal L}^{-1}_R e^{-2\rho} \partial^2 + 3
 {\cal L}^{-1}_R \delta(t - t_i) \partial_t + 
\bm{x} \cdot \partial_{\sbm{x}} \,,
\label{Def:calD}
\end{eqnarray}
where we used the facts 
\begin{eqnarray}
  {\cal L}^{-1}_R \zeta_I {\cal R}_x \zeta_I
 \Approx \zeta_I {\cal L}^{-1}_R {\cal R}_x \zeta_I\,, \label{Prop1}
\end{eqnarray}
and ${\cal L}^{-1}_R f(x) \Approx 0$ for $f(x) \Approx 0$~\cite{SRV1}.
In the above expression (\ref{Exp:gz2})
$\bar\zeta_I$ appears in the combination of $\bar\zeta_I {\cal D}_x$. 
To be more precise, the terms with the delta function $\delta(t-t_i)$ in ${\cal D}_x$ are
multiplied by $\bar{\zeta}_I(t_i)$, while the remaining terms are
multiplied by $\bar{\zeta}_I(t)$. Therefore, the former terms contribute
only to the IRdiv, while the latter terms contribute to both the
IRdiv and IRsec. First, we focus on the IRdiv, neglecting the IRsec
for a while. Inserting Eqs.~(\ref{Exp:gz2}) into Eq.~(\ref{Exp:1loop}), 
we obtain the one loop correction with the factor 
$\langle \bar{\zeta}^2_I(t_i) \rangle$ as 
\begin{eqnarray}
 &\langle {\cal R}_{x_1}\! \gz(x_1) {\cal R}_{x_2}\! \gz(x_2)
  \rangle_{1 {\rm loop}}  \nonumber \\
  &\Approx \frac{\langle \bar{\zeta}^2_I(t_i) \rangle}{2} {\cal R}_{x_1}  {\cal R}_{x_2} \Big\langle  2 {\cal
   D}_{x_1} \zeta_I(x_1){\cal D}_{x_2} \zeta_I(x_2) \cr
  & \qquad \qquad  \qquad \qquad  \qquad \quad   + {\cal
   D}_{x_1}^2 \zeta_I(x_1) \zeta_I(x_2)  + \zeta_I(x_1){\cal D}_{x_2}^2 \zeta_I(x_2)  
 \Big\rangle. \label{Exp:1loop2}
\end{eqnarray}
In general, the above expression is not free from the IRdiv and IRsec. 
One may think that the absence of the IRdiv requests ${\cal D}_x \zeta_I(x) =0$. 
However, this condition immediately contradicts, because an operation of
$\bm{x} \cdot \partial_{\sbm{x}}$ on a Fourier mode $e^{i\sbm{k}\cdot\sbm{x}}$ yields  
the term with $(\bm{x} \cdot \bm{k}) e^{i\sbm{k}\cdot\sbm{x}}$, which
cannot be canceled by the remaining two terms with the retarded integral
${\cal L}^{-1}_R$~\cite{SRV1}.

A simple alternative way we can think of is to impose
\begin{eqnarray}
 {\cal D}_x \zeta_I(x)=
\int{\dd^3 \bm{k} \over (2\pi)^{3/2}} \left(
  a_{\sbm{k}} D e^{i\sbm{k}\cdot\sbm{x}} v_k+ ({\rm h.c.}) \right)\, 
\label{bettercondition}
\end{eqnarray}
where 
\begin{eqnarray}
 & D \equiv k^{-3/2} e^{-i\phi(k)}
   \bm{k} \cdot \partial_{\sbm{k}} k^{3/2}e^{i\phi(k)}\,, 
\end{eqnarray}  
with an arbitrary phase function $\phi(k)$. (The choice of the
$\bm{k}$-dependent phase in the mode functions is irrelevant from the
beginning as usual.) With these conditions, the terms with
$\langle \bar{\zeta}_I^2(t_i) \rangle$ in Eq.~(\ref{Exp:1loop2}) 
can be summarized in the total derivative form as
\begin{eqnarray*}
  &\langle {\cal R}_{x_1}\! \gz(x_1) {\cal R}_{x_2}\! \gz(x_2)
  \rangle_{1 {\rm loop}} \nonumber \\
  &\,\, \Approx
\frac{ \langle \bar\zeta_I^2(t_i)  \rangle }{2} {\cal R}_{x_1} {\cal R}_{x_2}  \int \frac{\dd (\ln k) 
\dd \Omega_{\sbm{k}}
}{(2\pi)^3} \partial^2_{\ln k}
 \left\{ k^3 |v_k|^2 e^{i \sbm{k}\cdot (\sbm{x}_1-\sbm{x}_2)}\right\}\,,
\end{eqnarray*} 
where $\int \dd \Omega_{\sbm{k}}$ denotes the
integration over the angular directions of $\bm{k}$. 
Since the integral of a total derivative vanishes, 
the IRdiv is eliminated. 
The condition (\ref{bettercondition}) can be rewritten 
as a condition on mode functions 
\begin{equation}
  {\cal L}_{R,k}^{-1} \left( -2 (k e^{-\rho})^2 + 3
 \delta(t - t_i) \partial_t \right)  v_k=D v_k\,,
\label{Cond:GI}
\end{equation}
where ${\cal L}^{-1}_{R,k}$ is the Fourier mode of 
${\cal L}^{-1}_R$. 
We request the mode functions $v_k$ to satisfy Eq.~(\ref{Cond:GI}) and its time derivative just after 
the initial time $t=t_i$. Then, the condition (\ref{Cond:GI}) continues to
hold also for $t>t_i$, because this condition vanishes
under the operation of the second order differential operator 
${\cal L}$. 

Similarly, we can also discuss the absence of the IRsec, which
requests the terms with 
\begin{eqnarray*}
 \langle \bar{\zeta}_I^2(t) \rangle - \langle \bar{\zeta}_I^2(t_i)
\rangle \simeq \int_{1/L_{t_i} \leq k \leq 1/L_t} \frac{\dd^3 \bm{k}}{(2\pi)^3} P(k)
\end{eqnarray*}
should vanish. Recalling that $\bar\zeta_I$ associated with
$\delta(t-t_i)$ is to be interpreted as $\bar\zeta_I(t_i)$, we find that
the terms which contain $ \langle \bar{\zeta}_I^2(t) \rangle - \langle 
\bar{\zeta}_I^2(t_i) \rangle$ vanish as an integral of a total derivative, 
if 
\begin{equation}
  - 2{\cal L}_{R,k}^{-1} (k e^{-\rho})^2 v_k=D v_k\,
\label{Cond:GI2}
\end{equation}
is satisfied.

In this subsection we have observed that requesting the absence of
the IRdiv and IRsec restricts the initial states. 
In the succeeding subsection, we will show that the same
conditions as Eqs.~(\ref{Cond:GI}) and (\ref{Cond:GI2}) are derived from the requirement that the quantum state 
is invariant under the dilatation transformation, 
which will clarify the physical meaning of these conditions. 

\subsection{The canonical systems connected by the dilatation}
\label{SSec:CV}
To discuss the physical meaning of the 
IR regularity conditions, we introduce another
set of the canonical variables. As long as we consider a theory which
preserves the three-dimensional diffeomorphism invariance, 
the dilatation symmetry, 
$\bm{x} \to e^{-s} \bm{x}$ with a constant parameter $s$, 
is preserved as a part of spatial coordinate transformations. 
This implies that the action for $\zeta$ should be
invariant under the change of the variable
$\zeta(t,\bm{x}) \to  \zeta(t,\, e^{-s}\bm{x}) - s$, {\it i.e.},
\begin{eqnarray}
 & S = \int \dd t\, \dd^3\! {\bm x}\,  {\cal L}\left[\zeta(x) \right] = 
 \int \dd t\, \dd^3\! {\bm x}\, {\cal L}\left[ \zeta(t,\, e^{-s}\bm{x})
 - s \right]\,. \label{Exp:Dsym}
\end{eqnarray}

We introduce another set of
canonical variables than $\zeta(x)$ and its conjugate momentum $\pi(x)$
by  
\begin{eqnarray}
 & \tilde{\zeta}(x) \equiv \zeta(t,\, e^{-s}\bm{x}) \,, \qquad \quad
 \tilde{\pi}(x) \equiv  e^{-3 s}\pi(t,\, e^{-s}\bm{x})\,. \label{Def:tilde}
\end{eqnarray}
One can show that,
using the commutation relations for $\zeta(x)$ and $\pi(x)$, 
these new variables also 
satisfy 
the canonical commutation relations 
\begin{eqnarray}
 & \left[ \tilde{\zeta}(t,\, {\bm x}),\, \tilde{\pi}(t,\, {\bm y})
 \right] = \left[ \zeta(t,\, e^{-s}{\bm x}),\, e^{-3
 s} \pi(t,\, e^{-s}{\bm y})  \right]
 = i \delta^{(3)}({\bm x}- {\bm y})\,,
\end{eqnarray}
and 
\begin{eqnarray}
 &  \left[ \tilde{\zeta}(t,\, {\bm x}),\, \tilde{\zeta}(t,\, {\bm y})
 \right] =  \left[ \tilde{\pi}(t,\, {\bm x}),\, \tilde{\pi}(t,\, {\bm y})
 \right] = 0\,.
\end{eqnarray}

Using Eq.~(\ref{Exp:Dsym}), we can show that the Hamiltonian densities
expressed in terms of these two sets of the canonical variables are
related with each other as
\begin{eqnarray}
\! \int \dd^3\! {\bm x}\, {\cal H}[\zeta(x),\, \pi(x)] &= \int 
 \dd^3 {\bm x} \left\{ \pi(x) \dot\zeta(x) - {\cal L}[\zeta(x)] \right\} \cr
 & = \int \dd^3\! {\bm x}\, {\cal H}[\tilde{\zeta}(x)- s,\,
 \tilde{\pi}(x)] \cr
 & \equiv \int \dd^3\! {\bm x}\, \tilde{{\cal H}}[\tilde{\zeta}(x),\,
 \tilde{\pi}(x)], 
\label{Exp:Ht}
\end{eqnarray} 
where we changed the spatial coordinates 
as $\bm{x} \to e^{-s} \bm{x}$ on the second equality.
We find that the Hamiltonian density for the system $\cvt$ is
given by the same functional as the one for the system $\cv$ with
$\tilde{\zeta}$ shifted by $-s$.  

We can extend
the dilatation transformation to a time
dependent one, $s\to s(t)$ \cite{SRV2}. 
Again, we introduce another set of the canonical 
variables $\tilde{\zeta}(x)$ and $\tilde{\pi}(x)$,
using $\zeta$ and $\pi$ evaluated at the transformed point
$ e^{-s(t)}\bm{x}$,  
\begin{eqnarray}
 & \tilde{\zeta}(x) \equiv \zeta(t,\, e^{-s(t)}\bm{x}) \,, \qquad \quad
 \tilde{\pi}(x) \equiv  e^{-3 s(t)}\pi(t,\, e^{-s(t)}\bm{x})\,, \label{Def:tildet}
\end{eqnarray}
which satisfy the canonical commutation relations. The Hamiltonian
density for the variables $\{\tilde{\zeta},\, \tilde{\pi} \}$ can
be expressed in terms of the one for $\{\zeta,\, \pi \}$ as
\begin{eqnarray}
   \tilde{{\cal H}} \left[ \tilde{\zeta}(x),\, \tilde{\pi}(x) \right] 
    = {\cal H}  \left[ \tilde{\zeta}(x) - s(t),\, \tilde{\pi}(x) \right]  
  - \dot{s}(t) \tilde{\pi}(x) \bm{x} \cdot \partial_{\sbm{x}}
  \tilde{\zeta}(x)\,. \label{Exp:tH}
\end{eqnarray}

Assuming that $s(t)$ is as small as $\tilde{\zeta}(x)$ and $\tilde{\pi}(x)$, we
decompose the Hamiltonian densities ${\cal H}$ and $\tilde{\cal H}$ into
the quadratic free parts and the higher-order interaction parts as 
\begin{eqnarray}
 & {\cal H}[ \zeta(x),\,\pi(x)] = {\cal
 H}_0[\zeta(x),\, \pi(x)] +{\cal H}_I [\zeta(x),\,\pi(x)]\,,
\end{eqnarray}
and
\begin{eqnarray}
 & \tilde{\cal H} \left[ \tilde{\zeta}(x),\, \tilde{\pi}(x) \right] = {\cal
 H}_0 \left[ \tilde{\zeta}(x),\, \tilde{\pi}(x)  \right] +
 \tilde{{\cal H}}_I \left[ \tilde{\zeta}(x),\, \tilde{\pi}(x) \right]\,.  \label{Exp:Hdec}
\end{eqnarray}
Here, we note that 
${\cal H}_0 [ \tilde{\zeta}(x) -s(t),\, \tilde{\pi}(x)]=
  {\cal H}_0 [ \tilde{\zeta}(x),\, \tilde{\pi}(x)]$, 
since $\zeta(x)$ always 
appears with spatial differentiations in ${\cal H}_0 [\zeta(x),\pi(x)]$. 
Remarkably, the quadratic part of the 
Hamiltonian densities ${\cal H}$ and $\tilde{\cal H}$ have the same
functional form. Using Eq.~(\ref{Exp:tH}), we find that the interaction
Hamiltonian densities are related with each other as 
\begin{eqnarray}
 &   \tilde{{\cal H}}_I \left[\tilde{\zeta}(x),\, \tilde{\pi}(x) \right] 
 \equiv {\cal H}_I \left[ \tilde{\zeta}(x) - s(t),\, \tilde{\pi}(x)  \right]  -
 \dot{s}(t)  \tilde{\pi}(x)\, \bm{x}\! \cdot\! \partial_{\sbm{x}}
 \tilde{\zeta}(x) \,.  \label{Def:tH}
\end{eqnarray}
Thus, we find that $\tilde{\zeta}(x)$ in $\tilde{{\cal H}}_I$ only 
appears in the form of $\tilde{\zeta}(x)-s(t)$ or with 
differentiations. 

\subsection{The gauge invariance and the IR regularity}  \label{SSec:IRdivC}
Now, we are ready to give an alternative interpretation of the
conditions (\ref{Cond:GI}) and (\ref{Cond:GI2}). 
We adopted the initial condition (\ref{IC}) for the system $\cv$, 
which identifies the Heisenberg fields with the corresponding 
interaction picture fields at the initial time and 
selected the vacuum state at the initial time by (\ref{Def:vacuum}). 
These procedures specify a quantum state for the system $\cv$. 
If we adopt the same scheme in the canonical system $\cvt$, 
it is not obvious whether physically the same vacuum state 
is picked up or not. 
In this subsection, we show that 
these two vacua are equivalent 
only when the conditions (\ref{Cond:GI}) and (\ref{Cond:GI2}) are satisfied. 

Adopting the same scheme to give a quantum state, 
both $\zeta$ and $\tilde{\zeta}$ are solved by using
${\cal L}^{-1}_R$ with the conditions 
identifying the Heisenberg fields to the interaction picture fields 
at the initial time. 
We expand the respective interaction picture
fields, $\zeta_I$ and $\tilde{\zeta}_I$, 
in terms of the same mode function $v_k$ as
\begin{eqnarray}
 & \zeta_I(x) = \int \frac{\dd^3 \bm{k}}{(2\pi)^{3/2}} a_{\sbm{k}} v_k(t)
 e^{i \sbm{k}\cdot \sbm{x}} + ({\rm h.c.})\,,  \\
&  \tilde\zeta_I(x) = \int \frac{\dd^3 \bm{k}}{(2\pi)^{3/2}} \tilde{a}_{\sbm{k}} v_k(t)
 e^{i \sbm{k}\cdot \sbm{x}} + ({\rm h.c.})\,.   \label{Exp:tzI}
\end{eqnarray}
Then, we choose the vacuum states, $|\,0\, \rangle$ and 
$|\,\tilde{0}\, \rangle$ erased by operations of $a_{\sbm{k}}$ and
$\tilde{a}_{\sbm{k}}$, respectively. Now we compare the two-point functions calculated in 
the respective systems to show that the requirement 
\begin{eqnarray}
 & \langle\,0\,| \zeta(x_1) \zeta(x_2) |\,0\,\rangle 
 = \langle\,\tilde{0}\,| \tilde\zeta(t,\, e^{s(t)} \bm{x}_1)
 \tilde{\zeta}(t,\, e^{s(t)} \bm{x}_2) 
 |\,\tilde{0}\,\rangle \, \label{Exp:CT2}
\end{eqnarray}
yields the conditions (\ref{Cond:GI}) and (\ref{Cond:GI2}). 
We expand $\tilde\zeta(t,\, e^{s(t)} x^i)$ as
\begin{eqnarray}
 \tilde{\zeta}(t,\, e^{s(t)} \bm{x}) &=
 \tilde{\zeta}(x) + s(t) \bm{x} \cdot \partial_{\sbm{x}} \tilde{\zeta}(x) +
 \frac{1}{2} s^2(t) (\bm{x} \cdot \partial_{\sbm{x}})^2 \tilde{\zeta} (x) +
 {\cal O}(s^3)  \cr
 & \Approx \tilde{\zeta}_I(x) - \tilde\zeta_I (x) {\cal L}^{-1}_R \!  \left( 2
  e^{-2\rho} \partial^2 + 3 \delta(t - t_i) \partial_t \right)
 \tilde\zeta_I(x) \cr
 & \quad \quad + s(t) {\cal D}_x \tilde\zeta_I(x)
 + \cdots \,,
\end{eqnarray}
where ``$\cdots$'' denotes higher order terms in perturbation.
Note that the interaction Hamiltonian density for $\cvt$ is given by 
the same functional as ${\cal H}_I$ with the argument shifted by
$-s(t)$ and the second term in Eq.~(\ref{Exp:tH}), which is higher order
in perturbation. Then, the right-hand side of Eq.~(\ref{Exp:CT2}) gives
\begin{eqnarray}
 &  \langle\,\tilde{0}\,| \tilde\zeta(t,\, e^s \bm{x}_1) \tilde{\zeta}(t,\, e^s \bm{x}_2)
 |\,\tilde{0}\,\rangle  \cr
 &  \Approx  \langle\,0\,| \zeta(x_1) \zeta(x_2) |\,0\,\rangle   \cr
 & \quad  + s(t)  \langle\,\tilde{0}\,| {\cal D}_{x_1} \tilde\zeta_I (x_1) \tilde{\zeta}_I(x_2)\! + \cdots
 |\,\tilde{0}\,\rangle  +  s(t)  \langle\,\tilde{0}\,| \tilde\zeta_I
 (x_1) {\cal D}_{x_2} \tilde{\zeta}_I(x_2) \!+ \cdots  |\,\tilde{0}\,\rangle \cr
 & \quad   + {\cal O}(s^2)\,.  \label{Exp:CT3}
\end{eqnarray}
Now, it is clear
that Eq.~(\ref{Exp:CT2}) implies that the terms proportional to 
$s(t)$ on the right-hand side in Eq.~(\ref{Exp:CT3}) should vanish.

Notice that, more precisely, $s(t)$ multiplied by the term ${\cal L}^{-1}_R \delta(t- t_i)(\cdots)$ in ${\cal D}_x$
should be replaced with $s(t_i)$. Now we decompose $s(t)$ 
in Eq.~(\ref{Exp:CT3}) into $s(t_i)$ and $s(t)-s(t_i)$.
Then, we can see that if the condition for the absence of 
IRdiv~(\ref{Cond:GI}) is valid, the terms multiplied by $s(t_i)$ vanish. 
Whilst, if the condition for the absence of the IRsec~(\ref{Cond:GI2}) is valid, 
the terms multiplied by $s(t) - s(t_i)$ vanish. 
Thus, the conditions for the invariance under the 
dilatation~(\ref{Exp:CT2}) gives the same conditions 
as requesting the absence of the IRdiv and IRsec~\cite{SRV1}.

\subsection{Inconsistency in removing the IRdiv and IRsec}
\label{SSec:summary} 
In this section, we assumed that the
interaction is turned on at a finite initial time. However, 
we should also note that the conditions for the absence of 
the IRdiv and IRsec cannot be naturally satisfied 
in this setup. Regarding the condition for the absence of IRdiv 
(\ref{Cond:GI}), since the
left hand side of  (\ref{Cond:GI}) vanishes at $t=t_i$,
the condition (\ref{Cond:GI}) yields 
\begin{equation}
 D v_k(t_i)\simeq 0\,, \qquad D \dot{v}_k(t_i)=3\dot{v}_k(t_i)\,.
 \label{IRdivfree}
\end{equation} 
These conditions are compatible with the normalization 
of the mode functions 
\begin{equation}
-2i\Mp^2 e^{3\rho} \varepsilon_1 \left(
  v_k  \dot v^*_k
  -\dot v_k v_k^*\right)=1\,. 
\label{Eq:KGnormalization}
\end{equation}
Operating the derivative 
$\partial_{\ln k}$ on it, we find that the both sides vanish. 
However, the first condition in (\ref{IRdivfree})
requests the scale invariant spectrum with 
$|v_k(t_i)| \propto 1/k^{3/2}$ for all wavenumbers, which is not
compatible with the Hadamard condition in the UV limit. 
Therefore, it cannot be a physically natural quantum state. 

If we introduce an IR cutoff by hand, the IRdiv can be eliminated. 
Even in this case,  we cannot eliminate the IRsec. 
Since the left hand side of  (\ref{Cond:GI2}) and its time derivative vanish
at the initial time, the condition  (\ref{Cond:GI2}) requests
\begin{equation}
 D v_k(t_i)=0\,, \qquad D \dot{v}_k(t_i)=0\,, 
\end{equation}
which are incompatible with the normalization 
condition~(\ref{Eq:KGnormalization}). The right hand side
of (\ref{Eq:KGnormalization}) trivially vanishes after the operation 
of $\partial_{\ln k}$, while the left hand side gives 3.

The intuitive reason why these conditions cannot be compatible with the initial
condition (\ref{IC}) is as follows. When
we abruptly switch on the interaction at $t=t_i$, 
we introduce a particular time into the system. 
Then, the Hubble scale at
the initial time is distinguished from other scales. 
Therefore, the invariance under the dilatation transformation 
is naturally broken. 
One possible way to avoid this symmetry breaking 
might be sending the initial time to the infinite past. 
In this limit, 
the IR regularity no longer requests the condition (\ref{Cond:GI}),
because Eq.~(\ref{Prop1}) does not hold.

\section{IR regularity of the Euclidean vacuum in the inflationary universe}
\label{Sec:Euclidean}
In the previous section we showed that the \gauge invariance
in the local observable universe is essential to remove
the IRdiv and IRsec. However, we found it impossible to prepare a natural 
quantum state that maintains the \gauge invariance,  
as long as we start with the vacuum state of 
the free field, turning on the interaction at a finite time. 
In this section, we keep the interaction
turned on from the infinite past. We consider the Euclidean vacuum, which 
is specified by the regularity at the infinite
past with the time coordinate rotated towards the imaginary axis.  
As will be explained, the Euclidean vacuum keeps the \gauge
invariance, and hence the loop corrections become IR regular. 
We also show the absence of the SG without neglecting
the subH modes (to a certain order in the perturbative expansion). 

\subsection{The Euclidean vacuum}
In the case of a massive scalar field in de Sitter spacetime,
the boundary condition specified by rotating the time path in 
the complex plane can be understood as 
requesting the regularity of correlation functions on the Euclidean
sphere which can be obtained by the analytic continuation from those on
de Sitter spacetime (see Sec.~\ref{SSec:dSinv}). 
The vacuum state thus defined is called the Euclidean vacuum
state. Here, we also refer to the state which is
specified by a similar boundary condition in more general spacetime
as the Euclidean vacuum.

To be more precise, 
we define the Euclidean vacuum, requesting the regularity of the
$n$-point functions,  
\begin{eqnarray}
 &  \langle T_c\, \zeta(x_1) \cdots \zeta(x_n) \rangle < \infty\,, \qquad {\rm for}\,\,
  \quad \eta(t_a) \to  - \infty (1 \pm i \epsilon)\,,  \label{Def:EV}
\end{eqnarray}
where $a=1, \cdots, n$ and $T_c$ denotes the path ordering along the closed
time path, $-\infty(1-i\epsilon)\to \eta(t_f) \to -\infty(1+i\epsilon)$, in 
the conformal time defined in Eq.~(\ref{Def:eta}). For
simplicity, here we assume that 
$e^{\rho(t)} \dot\rho(t)$ is rapidly increasing in time so that
\begin{eqnarray}
  |\eta(t)|={\cal O} \left( 1/e^{\rho(t)} \dot{\rho}(t) \right)\,.
 \label{eta}
\end{eqnarray}

The Euclidean vacuum 
is expected to possess the \gauge invariance in the local universe, 
especially the invariance under the dilatation transformation,  
since its conditions do not introduce any artificial scale. 
In fact, the Euclidean vacuum is specified independently of
which canonical variables $\cv$ or $\cvt$ we use. 
The boundary conditions of the Euclidean vacuum for the
canonical variable $\tilde{\zeta}$ request 
\begin{eqnarray}
 &  \langle T_c\, \tilde\zeta(x_1) \cdots \tilde\zeta(x_n) \rangle < \infty\,, \qquad {\rm for}\,\,
  \quad \eta(t_a) \to  - \infty (1 \pm i \epsilon)\,.  \label{Def:EVt}
\end{eqnarray}
Then, we can show the equivalence
\begin{eqnarray}
 & \langle T_c\, \zeta(x_1) \cdots \zeta(x_n)
 \rangle = \langle T_c \tilde{\zeta}(t_1,\,
 e^{s(t_1)}\bm{x}_1)  \cdots \tilde\zeta(t_n,\, e^{s(t_n)} \bm{x}_n)
 \rangle \label{Exp:EV}
\end{eqnarray}
is satisfied. This is a generalization of the condition (\ref{Exp:CT2}),
which requests the invariance under the dilatation transformation. 
A more detailed explanation
regarding the uniqueness of the Euclidean vacuum can be found in
Ref.~\cite{SRV2}. The distinctive property~(\ref{Exp:EV}) 
will be the key to show that 
the Euclidean vacuum is free from the IRdiv and IRsec.

Here we took the boundary conditions for the $n$-point functions 
as the definition of the Euclidean vacuum state, assuming the existence of such a quantum state. 
In the in-in formalism, the $n$-point functions are perturbatively 
expanded using the Wightman functions. 
At this point, the vertex integrals along the closed time path 
start and end with ${\rm Re}[\eta] \to - \infty$.
The infinitely oscillating vertex integrals along this path 
can be made convergent by rotating the time path toward the
imaginary axis, which is nothing but the ordinary $i\epsilon$
prescription. Thus obtained $n$-point functions also satisfy the boundary conditions 
 (\ref{Def:EV})/(\ref{Def:EVt}) (see Sec. IV A of Ref.~\cite{SRV2}). 

\subsection{The IR regular Hamiltonian}
In this subsection, we discuss the quantization using the
canonical variables $\cvt$. 
When we choose the Euclidean vacuum, the interaction Hamiltonian density
for $\cvt$, $\tilde{\cal H}_I$ can be recast into the form  
\begin{eqnarray}
 & \tilde{\cal H}_I[\tilde{\zeta}_I(x),\, \tilde{\pi}_I(x)] = \Mp^2
  e^{3\rho} \dot{\rho}^2 \varepsilon_1(t) \sum_{n=3}^\infty \lambda(t) 
 \prod_{m=1}^n {\cal R}^{(m)}_x \tilde{\zeta}_I(x) \label{Exp:tHIRR}
 \,,
\end{eqnarray}
where $\lambda(t)$ represents an ${\cal O}(1)$ dimensionless time-dependent 
function expressed in terms of the horizon flow functions. 
To discriminate different IR suppressing operators, we accompany ${\cal R}_x$ with a superscript $(m)$.  
Although the interaction Hamiltonian for the curvature
perturbation is very messy, what we need to verify Eq.~(\ref{Exp:tHIRR}) is
only the formal expression given in Eq.~(\ref{Exp:tH}), {\it i.e.}, 
$\tilde{{\cal H}}_I$ can be written down solely in terms of 
$\tilde{\zeta}_I(x)-s(t)$, $\tilde{\zeta}_I(x)$ with differentiation, 
and $\dot{s}(t)$ as a manifestation of the dilatation symmetry. 
We should note that
the expression~(\ref{Exp:tH}) does not immediately imply 
that $\tilde{{\cal H}}_I$
is composed of IR irrelevant operators because of the
terms with $\tilde{\zeta}_I(x)-s(t)$ and the inverse Laplacian
$\partial^{-2}$.  

First, we consider the terms with $\tilde{\zeta}_I(x)-s(t)$. 
Owing to the
uniqueness of the Euclidean vacuum discussed in the previous
subsection, we can replace all $\{\tilde{\zeta}_I(x)-s(t)\}$s in the
interaction Hamiltonian with $\tilde{\zeta}_I(x)-\Gbz_I(t)$, which is in
the form ${\cal R}_x\tilde{\zeta}_I(x)$~\cite{SRV2}. 
This replacement introduces
additional terms, but these terms are shown to be products of
$\tilde{\zeta}_I(x)$s suppressed by ${\cal R}_x$. Similarly, we can replace all
$\dot{s}(t)$s with the terms which are products of 
${\cal R}_x \tilde{\zeta}_I(x)$. Thus, we can show that all the
interaction picture fields $\tilde{\zeta}_I(x)$ are multiplied by the IR
suppressing operator ${\cal R}_x$. 

Next, we consider the inverse Laplacian
$\partial^{-2}$, which appears in solving the constraint equations 
for the lapse function and the shift vector. If $\partial^{-2}$
does not introduce additional inverse power of $k$,  
Eq.~(\ref{Exp:tHIRR}) is verified. Repeating the
discussion about the boundary condition 
of $\partial^{-2}$ in Sec.~\ref{SSSec:GF}, we can
restrict all the interaction vertexes within the causally connected
local region ${\cal O}$, prohibiting the appearance of additional
inverse power of $k$. When we calculate $n$-point functions for the genuinely
\gauge invariant operator $\gR$ from those for ${\cal R}_x \gz(x)$, 
the choice of the boundary
conditions should not affect the results. 
In this way, we can express all the
interaction Hamiltonian in the form of Eq.~(\ref{Exp:tHIRR}). 

\subsection{The regularized Wightman function}  \label{SSec:Wightman}
Since all $\tilde{\zeta}_I(x)$s in the interaction
Hamiltonian are multiplied by the IR suppressing operators 
${\cal R}_x$, the $n$-point function of ${\cal R}_x \gz(x)$ can be
expanded by the Wightman function 
${\cal R}_x {\cal R}_{x'} G^+(x,\,x')$ and its complex conjugate
${\cal R}_x {\cal R}_{x'} G^-(x,\, x')$. 
In this subsection, we calculate these Wightman functions multiplied by
the IR suppressing operators. We will find that the boundary condition of the
Euclidean vacuum guarantees that the amplitude of 
${\cal R}_x {\cal R}_{x'} G^+(x,\, x')$ is bounded from above 
for finite values of $x$ and $x'$, except for the coincidence limit.

As mentioned above, the boundary conditions of the Euclidean vacuum
(\ref{Def:EV})/(\ref{Def:EVt}) are equivalent to 
the $i\epsilon$ prescription in the in-in formalism. 
We expand the curvature perturbation $\tilde{\zeta}_I(x)$ as in
Eq.~(\ref{Exp:tzI}), using the mode function $v_k(t)$. The
boundary conditions (\ref{Def:EV})/(\ref{Def:EVt}) at the
tree level imply that 
$v_k(t)$ should be $\propto e^{-ik\eta(t)}$ asymptotically. 
Factoring out this time dependence, we express $v_k(t)$ as 
\begin{eqnarray}
 & v_k(t) = 
\frac{{\cal A}(t)}{k^{3/2}} f_k(t) e^{-i k\eta(t)}\,, \label{Exp:vk}
\end{eqnarray}
where ${\cal A}(t)$ is an approximate amplitude of the fluctuation 
defined by
\begin{eqnarray}
  {\cal A}(t) \equiv \frac{\dot{\rho}(t)}{\sqrt{\varepsilon_1(t)} \Mp}\,.
\end{eqnarray}
Using Eq.~(\ref{Exp:vk}) and integrating over the angular part of the
momentum, the Wightman function ${\cal R}_x {\cal R}_{x'} G^+(x,\,x')$ 
can be expressed as
\begin{eqnarray}
 {\cal R}_x {\cal R}_{x'} G^+(x,\, x')
 &= {1 \over 2\pi^2}
  \int^{\infty}_0 \frac{\dd
  k}{k}\,  {\cal R}_x{\cal R}_{x'} {\cal A}(t) f_k(t)
{\cal A}(t') f_k^*(t') \cr
 & \qquad \quad  \times 
\Biggl[{e^{i k \sigma_+(x, x')} -e^{i k \sigma_-(x, x')} \over
  i k (\sigma_+(x, x') -  \sigma_-(x, x'))}
\Biggr]\,, 
\label{Exp:DG+}
\end{eqnarray}
where we introduced
$
\sigma_{\pm}  (x,\, x') \equiv \eta(t')-\eta(t) \pm  |\bm{x} - \bm{x}'|\,. 
$
The function $f_k(t)$ satisfies a regular second order differential 
equation with the regular boundary condition  
$f_k(t)\to k/(\sqrt{2}\,e^\rho \dot\rho)$ for $-k\eta(t) \to \infty$.
Since both the differential equation and the boundary condition for $f_k(t)$
are analytic in $k$, the resulting function $f_k(t)$ should be
analytic as well. Namely, $f_k(t)$ does not have any singularity
such as a pole on the complex $k$-plane. 

Now we are ready to discuss the regularity of
${\cal R}_x {\cal R}_{x'} G^+(x,\, x')$, particularly the regularity of the
$k$ integration in Eq.~(\ref{Exp:DG+}). Since the function $f_k(t)$ is
not singular, the regularity can be verified if the integration
converges both in the IR
and UV limits. The regularity in the IR limit is guaranteed by 
the IR suppressing operators ${\cal R}_x$, which add at least 
one extra factor of $k|\eta(t)|$ or eliminate the leading
$t$-independent term in the IR limit. In the UV limit, 
the contour of the $k$-integral in Eq.~(\ref{Exp:DG+}) should be
appropriately modified at $k\to\infty$ so that the integral becomes
convergent,  
which is a part of the $i\epsilon$ prescription. With this prescription, 
the integral is made finite for the UV contribution except for the case 
$\sigma_{\pm}(x, x') = 0$, where 
$x$ and $x'$ are mutually light-like. Since the expression of the
Wightman function obtained after the $k$ integration is independent of
the value of $\epsilon$, the regulator makes the UV contributions
convergent even after $\epsilon$ is sent to zero. 
For $\sigma_{\pm}(x, x') = 0$, 
the integral becomes divergent in the limit $\epsilon\to 0$, 
but this divergence is to be interpreted as the ordinary 
UV divergences, whose contribution to the vertex integrals 
must be renormalized by 
introducing local counter terms.

\subsection{The secular growth (SG)}
\label{SSec:regularity}
Since the amplitude of ${\cal R}_x {\cal R}_{x'} G^+(x,\, x')$ is shown
to be finite, 
we can verify the regularity of the $n$-point functions 
if the non-vanishing support of the integrands of the vertex
integrals is effectively restricted to a finite spacetime region. 
Since the interaction vertexes are restricted to 
the causal past with the appropriate choice of the residual \gauge
degrees of freedom, the question to address is whether the contribution 
from the vertexes in the distant past is effectively shut off or not. 
In this subsection, focusing on the long-term correlation, 
we will give an intuitive explanation why 
the vertex integrals of the $n$-point functions converge 
for the Euclidean vacuum (see Ref.~\cite{SRV2} for details). 

When we choose the Euclidean vacuum as the initial state, 
the deep IR modes $k|\eta| \ll 1$ 
are suppressed by the operation of ${\cal R}_x$ and the UV modes
$k|\eta| \gg 1$ are suppressed owing to the boundary
condition of the $i \epsilon$ prescription. 
Thus, only the modes around the Hubble scale, {\it i.e.},  
$k|\eta| \simeq k/e^{\rho} \dot{\rho}  = {\cal O}(1)$, remain 
to be relevant. 
Then, the Wightman function ${\cal R}_x {\cal R}_{x'} G^+(x,\, x')$ 
is necessarily suppressed when $\eta(t)/\eta(t') \ll 1$~\footnote{An 
explicit computation shows that in the limit $|\eta(t)| \ll |\eta(t')|$,
the Wightman function ${\cal R}_x {\cal R}_{x'} G^+(x,\, x')$ is suppressed as 
\begin{eqnarray}
  {\cal R}_x {\cal R}_{x'} G^+(x,\, x') 
  = {\cal A}(t) {\cal A}(t') \,
 {\cO}\!\left(\left( |\eta(t)| \over |\eta(t')| \right)^{
 {n_s+1 \over 2}}\right)
\label{OoM},
\end{eqnarray}
where $n_s$ is the spectral index.},
because, if $x$ and $x'$
are largely separated in time, any Fourier mode 
in the Wightman function cannot be at the Hubble scale 
simultaneously at $t$ and $t'$. 
When we consider the contribution of vertexes located in the distant
past,  
at least one Wightman function should 
satisfy $\eta(t)/\eta(t') \ll 1$, and therefore 
it is suppressed. When all the time integrations converge, being
dominated by the contributions at around $t=t_f$, 
we have an estimate,  
\begin{eqnarray}
 & \langle\,0| {\cal R}_{x_1} \gz(t_f,\, \bm{x}_1) \cdots  {\cal R}_{x_n}  \gz(t_f,\,
 \bm{x}_n) |0\, \rangle  ={\cal O}( \hat{\lambda}(t_f) \{ {\cal A}(t_f) \}^{{\cal N}})\,,
\end{eqnarray} 
where ${\cal N}\equiv N_f - 2N_v$ with 
$N_f$ and $N_v$ being the numbers of
$\tilde{\zeta}_I$s and the vertexes 
contained in the corresponding diagram, respectively.  

When we consider a diagram for which a cluster of vertexes in the
distant past is connected to the vertexes around the observation time by a single 
propagator, the IR suppression comes only from this propagator. 
If the past cluster of vertexes includes a sufficiently large number of
operators, the increase of the amplitude of fluctuation ${\cal A}(t)$ toward the 
past high energy regime may overtake the suppression due to this
propagator. This happens only when ${\cal N}$ is extremely large such as 
$1/\varepsilon_1 \simeq {\cal O}(10^2)$. We should also stress that the
SG is totally suppressed in the slow roll limit.

In Refs.~\cite{SZ1210, ABG}, the absence of the secular
growth is claimed by computing $\dot{\zeta}_{\sbm{k}}$ in the limit
$k/(e^\rho \dot{\rho}) \to 0$. In these papers, the mode
function in de Sitter spacetime, whose amplitude at large scales is given by a constant
Hubble parameter, is used in proving the conservation of the curvature
perturbation, while the time variation of the amplitude cannot be
neglected namely for the tIR modes. This leads to the quantitative discrepancy in the
evaluation of the SG from the one given above. For instance, in
Ref.~\cite{SZ1210}, ensuring that $\dot{\tilde{\zeta}}^{(n)}_L(x, t)$ given in Eq.~(22) of the
paper does not have long-term correlations is crucial in their proof. However, the locality 
is not necessarily valid, once we take into account the fact that in the
chaotic inflation, the amplitude of the fluctuation becomes larger and larger in the distant
past as $\dot{\rho} \propto e^{- \int \dd \rho \varepsilon_1}$. When we
neglect this effect by setting ${\cal A} \propto (\dot{\rho}/\sqrt{\varepsilon_1})$ to
constant, the above discussion will also lead to the
absence of the SG irrespective of the order of perturbation. Therefore the result here does not contradict the
conservation of the curvature perturbation they claimed.

Here we also comment on the related works~\cite{BGHNT10, GHT11, GS10}.
In these references, the authors showed that
the two point function which
contains the logarithmic IRdiv is related to the one which does not by the 
dilatation transformation (see Sec.\ref{SSec:LA}). In our terminology, the former
is $\langle \zeta(x) \zeta(x') \rangle$ and the later is $\langle \gz(x) \gz(x') \rangle$. 
Note that $\langle \gz(x) \gz(x') \rangle$ can still suffer from the
SG, which can be eliminated only for a limited class of quantum state that is invariant 
under the dilatation transformation. In fact, explicit realizations of such quantum
states that we know are limited to the Euclidean vacuum and its
variants.

\subsection{The summary of the proof}
Now we conclude that, when we choose the Euclidean vacuum, 
the $n$-point functions for
the genuinely \gauge invariant curvature perturbation contain neither
the IRdiv nor the IRsec. Furthermore, they do not suffer from the SG unless a very
high order in the perturbative expansion is concerned. (The outline of
the proof is depicted in Fig.~3 of Ref.~\cite{SRV2}.) We repeat two key
points which ensure the absence of the IRdiv, IRsec, and SG: 
\begin{itemize}
 \item Evaluating a genuinely \gauge invariant operator.
 \item Choosing the quantum state to be invariant under the dilatation
       transformation. 
\end{itemize}
The genuinely \gauge invariant operator should be entirely composed of the
Heisenberg picture field $\tilde{\zeta}(x)$ with ${\cal R}_x$, but the
operator is not necessarily expanded solely in terms of ${\cal R}_x \tilde{\zeta}_I(x)$. 
Therefore, even if we consider the correlators of genuinely \gauge
invariant operators, they can suffer from the IRdiv and IRsec. 
Then, the second point becomes important. 
By choosing the Euclidean vacuum, which is 
invariant under the dilatation transformation, 
the correlators of genuinely \gauge invariant operators
can be expanded only in terms of ${\cal R}_x \tilde{\zeta}_I(x)$, 
and thus the IRdiv and IRsec are eliminated.  
As we described in Sec.~\ref{SSec:Vdiv}, all IR and tIR modes 
are initially subH modes, and hence it is not satisfactory to neglect
subH modes from the beginning namely in examining the SG. 
When we choose the Euclidean vacuum, the UV
modes much below the Hubble length scale 
are also suppressed, leaving aside the ordinary UV divergences to 
be renormalized by the local counter terms.

\section{The IR issues in the absence of the gravitational fluctuation} \label{Sec:test}
So far, we discussed the fluctuation of the inflaton taking into 
account (the longitudinal mode of) the gravitational perturbation. Then,
the modes far beyond the Hubble scale are almost indistinguishable 
from the residual \gauge degrees of freedom, which is 
crucial in showing the absence of the IRdiv, IRsec and SG. 
By contrast, the
same argument does not apply to the IR issues of a test field in
a fixed quasi de Sitter background spacetime, which is frequently 
discussed as an approximate toy model to discuss the iso-curvature 
perturbations. 
In this section, we briefly summarize
the recent progress in this subject. 

\subsection{Resummation and the dynamical mass
  generation}\label{SSec:Mass}
As is described in Sec.~\ref{SSec:Vdiv}, the logarithmic IRdiv
originates from the scale invariant spectrum of a light field in an 
inflationary spacetime. It has been pointed out that resumming loop
diagrams leads to a dynamical mass generation, which will remedy the 
singular behaviour in the IR. Here, we consider a scalar field $\Phi$
with the quartic coupling $\lambda \Phi^4$ in a fixed background 
inflationary spacetime. In this section, we use the dimensionful 
scalar field $\Phi$ instead of the dimensionless scalar field $\phi$. 

\subsubsection{The stochastic approach}  \label{SSSec:stochastic}
The stochastic approach, initiated by Starobinsky~\cite{S}, describes
the evolution of the superH modes, $\Phi_{\rm sp}$, defined 
by eliminating the contribution from the subH modes as 
\begin{eqnarray}
  \Phi_{\rm sp}(x) = 
\Phi(x) -\int \frac{\dd^3
   \bm{k}}{(2\pi)^{3/2}} \theta(k-\epsilon e^{\rho(t)} \dot{\rho}(t))
 \left[ a_{\sbm{k}} \Phi_k(t) e^{i \sbm{k} \cdot \sbm{x}} + ({\rm h.c.}) \right]\,,
\end{eqnarray}
with a small positive parameter $\epsilon$. 
The evolution equation for $\Phi_{\rm sp}$ is given by  
\begin{eqnarray}
 \dot{\Phi}_{\rm sp}(x) = - \frac{1}{3 \dot{\rho}} \frac{\dd
  V(\Phi_{\rm sp})}{\dd \Phi_{\rm sp}} + f \,, \label{Eq:st}
\end{eqnarray}
in the slow roll approximation, where $f(x)$ denotes the stochastic
noise due to the modes with $k \approx \epsilon e^{\rho(t)} \dot{\rho}(t)$ 
whose variance is given by 
\begin{eqnarray}
 \langle f(x_1) f(x_2) \rangle 
 = \frac{\dot{\rho}^3}{4 \pi^2} \delta(t_1 - t_2)
 \frac{\sin \epsilon e^{\rho(t)} \dot{\rho}(t) |\bm{x}_1 - \bm{x}_2|}{\epsilon
 e^{\rho(t)} \dot{\rho}(t) |\bm{x}_1 - \bm{x}_2|} \,. \label{Eq:st2}
\end{eqnarray}
The Fokker-Planck equation obtained from
Eqs.~(\ref{Eq:st}) and (\ref{Eq:st2}) gives the probability distribution
function (PDF) for $\Phi_{\rm sp}$. (In Ref.~\cite{Riotto:2011sf}, the PDF
for the curvature perturbation was discussed based on the stochastic
approach.)   

Solving the Fokker-Planck equation,
Starobinsky and Yokoyama~\cite{SY} showed that the variance 
of a massless scalar field with a quartic potential
$\lambda \Phi^4$ approaches a constant value 
\begin{eqnarray}
 \langle \Phi_{\rm sp}^2 \rangle \to 
\frac{c}{\sqrt{\lambda}} \dot{\rho}^2\,,  \label{Exp:variance}
\end{eqnarray}
at late times, where $c$ is an ${\cal O}(1)$ numerical factor. Note
that the late time behaviour of the variance does not suffer
from the logarithmic enhancement. 
This equilibrium state can be understood as the balance
between the potential force (the first term in Eq.~(\ref{Eq:st})) 
and the quantum fluctuation (the second term). 

\subsubsection{The two-particle irreducible formalism}  \label{SSSec:2PI}
Riotto and Sloth pointed out that the late time behaviour of 
$\langle \Phi_{\rm sp}^2 \rangle$ signals the dynamical mass generation due
to the higher loop contributions~\cite{Riotto:2008mv}. 
Using the two-particle irreducible (2PI)
effective action, we can obtain the equation of motion for the resummed
propagator $G(x_1,\, x_2)$ (see Ref.~\cite{Berges:2004yj} and the
references therein). A local contribution in the self-energy
$\Sigma(x_1,\, x_2)$, which is proportional to the delta function
$\delta(x_1 - x_2)$, shifts the effective mass which appears in the
equation of motion for $G(x_1,\, x_2)$. (For instance, the left diagram
of Fg.~\ref{Fg:loops} gives a local contribution to the self-energy.)
Namely, for a quartic interaction, the dynamically generated mass is
given by $M^2_{\rm dyn} \sim \lambda G(x,\, x)$ at the leading order of
the 2PI expansion. Replacing $G(x,\, x)$ with the variance
(\ref{Exp:variance}), Riotto and Sloth claimed that resumming 
higher loops yields
the effective mass of order of the Hubble scale as
$M^2_{\rm dyn} \sim \sqrt{\lambda}\dot{\rho}^2$. Garbrecht and
Rigopoulos~\cite{Garbrecht:2011gu}, Serreau~\cite{Serreau11, Serreau12,
Serreau13}, and Arai~\cite{Arai11, Arai12, Arai13} discussed the dynamical mass generation in more
detail, using the 2PI formalism. In particular, in Ref.~\cite{Serreau12}
and Refs.~\cite{Arai12, Arai13}, they elaborated on the
UV contributions, identifying the necessary counter terms.

\subsubsection{The dynamical renormalization group}  
The mass generation due to the resummation was reported also by
Burges {\it et al}.~\cite{Burgess09, Burgess10}, based on the dynamical renormalization group (RG)
technique, which is useful to address the RG flow in the presence of the
secular growth in time. They focused on the secular growth which appears 
from the naive estimation of the accumulated superH modes. 
By using the conventional cosmological perturbation theory, the power
spectrum of the test field in the de Sitter space with the loop
correction is given by  
\begin{eqnarray}
  P_\Phi (k,\, \eta) = \frac{\dot{\rho}^2}{2 k^3} 
\left[ 1 + \delta  \ln  \left( k \over e^\rho \dot{\rho}  \right)
\right] \label{Eq:PPhi0}
\end{eqnarray}
with
\begin{eqnarray}
 \delta \equiv  \frac{\lambda}{3} \frac{\langle \Phi^2 \rangle}{\dot{\rho}^2}\,,
\end{eqnarray}
where the logarithmic term $\ln (k/\dot{\rho} e^\rho)$ appeared by integrating the
superH modes in time. Here we neglected the subH
contribution. The dynamical RG technique suggests resumming the loop correction as
\begin{eqnarray}
  P_\Phi (k,\, \eta) = \frac{\dot{\rho}^2}{2 k^3}
 \left( k \over e^\rho \dot{\rho}  \right)^\delta 
\left[ 1 + {\cal O} (\delta^2)  \right]\,.  \label{Eq:PPhi}
\end{eqnarray}
Since the power spectrum for the massive scalar field is given by
Eq.~(\ref{Eq:PPhi}) with $\delta = 2 M^2/(3\dot{\rho}^2)$, we see that the
resummation generates the mass of order
$M^2_{\rm res} \sim  \dot{\rho}^2 \delta \sim \lambda \langle \Phi^2
\rangle$, reproducing the result obtained by Riotto and
Sloth~\cite{Riotto:2008mv}. (An impact of the resummation was explored based on
different approaches in Refs.~\cite{Jatkar:2011ju,
Frob:2013ht}.)

The dynamical mass generation, addressed in this subsection, can be
interpreted as the thermalization process. Since the de Sitter space does
not possess a global timelike Killing vector, there is no positive
definite conserved energy, and the background spacetime plays the
role of the heat bath with the de Sitter temperature $\approx \dot\rho$.
A related particle decay process in the de Sitter
space was also discussed by Bros {\it et al.} in Refs.~\cite{Bros:2006gs,
Bros:2008sq, Bros:2009bz}.

\subsection{The regularity for the Euclidean vacuum}  \label{SSec:dSinv}
For the non-interacting theory, a systematic study of the de
Sitter invariant Euclidean vacuum was done by Mottola in
Ref.~\cite{Mottola:1984ar} and Allen in Ref.~\cite{Allen:1985ux}. Recently, 
Hollands~\cite{SH} and Marolf and Morrison~\cite{MM10, MMall, MM11}
systematically 
investigated the higher order loop corrections of a massive test scalar
field in the exact de Sitter space. They showed the perturbative stability of 
the Euclidean vacuum, which is identical to the so-called 
Bunch-Davies vacuum in the exact de Sitter case. 

The metric of the $D$-dimensional de Sitter space in global coordinates is given by 
\begin{eqnarray}
  \dd s^2 = - \dd t^2 + l^2 \left(\cosh \frac{t}{l}\right)^2 \dd
   \Omega_{D-1}^2\,, \label{dSD}
\end{eqnarray}
where $l\equiv 1/\dot{\rho}$ is the de Sitter length scale and 
$\dd \Omega_{D-1}^2$ is the metric on a unit $(D-1)$-dimensional sphere. 
Performing the Wick rotation from $t$ to 
$\tau \equiv \pi/2 - i (t/l)$, the metric (\ref{dSD}) is analytically
continued to the metric on a $D$-dimensional sphere, 
\begin{eqnarray}
  \dd s^2 = l^2 \left[ \dd \tau^2 + (\sin \tau)^2   \dd
   \Omega_{D-1}^2 \right]\,.
\end{eqnarray} 
It is obvious that the IRdiv is absent 
in the vertex integral over a compact manifold 
as long as the free propagator is well behaved there. Therefore, they first computed the $n$-point functions on the 
Euclidean sphere and then analytically continued the 
results to the Lorentzian domain. 

The de Sitter space can be described as the $D$-dimensional
hyperboloid with $\eta_{AB} X^A X^B= l^2$ embedded in the 
$(D+1)$-dimensional Minkowski spacetime with the metric tensor 
$\eta_{AB}$. The free propagator 
$\langle \Phi(x_1) \Phi(x_2) \rangle$ for the Bunch-Davies vacuum is
described only in terms of the invariant distance 
$Z_{12} \equiv  \eta_{AB} X^A(x_1) X^B(x_2)/l^2$. 
In Refs.~\cite{SH} and \cite{MM11} 
the $n$-point functions in the limit where two arguments are largely separated as 
$|Z_{12}| \gg 1$ was examined, and they showed that 
the loop corrections decays with the power in $|Z_{12}|$ not slower 
than the free field two-point function. The equivalence
between the correlators obtained after the analytic continuation 
and those computed in the Poicar$\acute{\rm e}$
patch, {\it i.e.}, the expanding cosmological patch, with the 
$i\epsilon$ prescription is shown for
interacting massive fields by Higuchi {\it et al}.~\cite{Higuchi:2010xt}
(and also by Korai and Tanaka in a different way~\cite{Korai:2012fi}).

By contrast, for a massless scalar field, the IR
regularity has not been shown and the absence of the SG
is unclear~\cite{KK10, KK1012, KK11, SH11}. 
Although the dynamical mass generation is one possible 
answer~\cite{Rajaraman:2010xd, Beneke:2012kn}, 
we can think of several possible situations in which 
mass generation is prohibited for the symmetry reason. 
The adiabatic curvature 
perturbation is a sort of
massless field whose mass generation is not allowed because 
of the diffeomorphism invariance. 
In this case the IR suppressing operators ${\cal R}_x$ 
are associated with the observable quantities 
by virtue of the residual \gauge symmetry.  
Because of that, although its Wightman function $G^+(x,\, x')$ 
behaves similar to a massless scalar field, 
the singular behaviour is 
cured for the Euclidean vacuum. 
In this sense, it would be intriguing to discuss a massless 
field with the exact shift symmetry in the de Sitter space (see also Ref.~\cite{Page}).

\subsection{The quantum decoherence and the IR problem}
Even if careful computations of correlation functions 
for observable quantities may 
give divergent results, 
this does not immediately indicate a pathology. 
This is because what we actually observe 
still can be different from what we compute 
based on the standard quantum field theory. 
The primordial perturbations are
supposed to decohere through the cosmic expansion and/or through various
interactions~\cite{S, Polarski:1995jg, Kiefer:2006je}. 
This decoherence process transmutes the quantum fluctuations 
at a long wavelength to a statistical ensemble. 
In the standard computation, however, the effect of 
this quantum decoherence is not taken into account, and hence 
the correlation functions that we calculate 
are the expectation values for a superposition of various 
wave packets which will never be observed simultaneously in reality.

\begin{figure}
\begin{center}
\includegraphics[width=15cm]{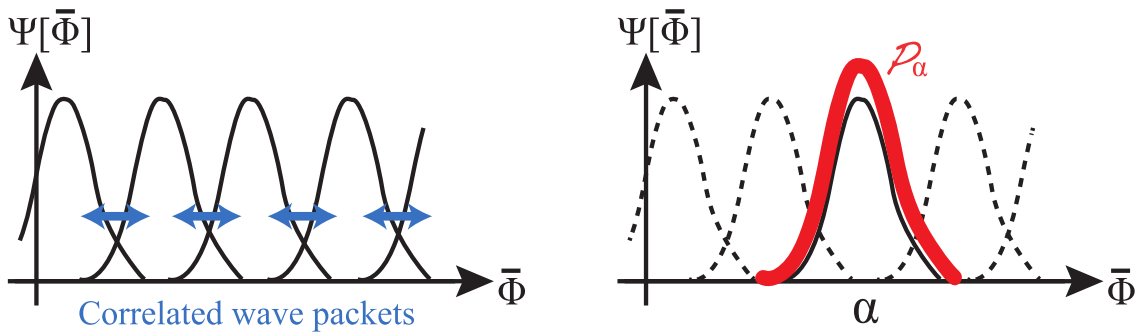}
\caption{The left panel shows the wave packets at the early stage of
 inflation, which are correlated with each other. At later times, the
 wave packets get decohered as depicted in the right panel. At observation, only one of the decohered
 wave packets is picked up. To eliminate the influence from the
 irrelevant wave packets to us, we introduce the operator ${\cal P}_\alpha$. }  
\label{Fg:dec}
\end{center}
\end{figure}

Here we focus on the spatial average of a test scalar field $\Phi$
given by 
$$
\bar{\Phi}(t)\equiv \frac{\int \dd^3\bm{x} W_t(\bm{x}) \Phi(t,\,\bm{x})}{\int
\dd^3\bm{x} W_t(\bm{x})}. $$  
When we choose an initial state with the scale invariant spectrum in the IR limit, the variance of
$\bar{\Phi}(t)$, {\it i.e.}, $\langle \bar{\Phi}(t)^2 \rangle$
diverges. The unbounded variance implies that the wave
function of $\bar{\Phi}$, $\Psi[\bar{\Phi}]$, 
does not have a sharp peak around a specific
value but spreads infinitely. 
As is shown in Fig.~\ref{Fg:dec}, such a wave function can be 
decomposed into a superposition of wave packets.  
Even if the quantum state starts with 
a coherent superposition of wave packets, 
the quantum coherence is gradually lost through the time evolution. 
Thus, at a later observation time $t_{\rm obs}$ 
the quantum coherence will be kept only among adjacent 
wave packets. Our universe will select a particular value 
$\bar{\Phi}(t_f)=\alpha$ once it is observed. 
This corresponds to picking up a single decohered
wave packet peaked at $\bar{\Phi}(t_f)=\alpha$ from the
superposition of wave packets. After the decoherence takes place, 
the other wave packets far from $\bar{\Phi}(t_f)\approx \alpha$ 
never contribute to observable quantities. 
Therefore it is more appropriate to remove the influence of these wave packets from the computation of observable quantities.

To take into account this selection effect, in Ref.~\cite{IRmulti}, we
proposed to insert an operator: 
\begin{eqnarray}
  {\cal P}_\alpha \equiv \exp \left[ - \frac{\{\bar{\Phi}(t_f) -
			       \alpha\}^2}{2 \sigma^2 \Mp^2} \right] \,.
\end{eqnarray} 
Here, $\sigma$ denotes the width of the projection window,  
which is supposed to satisfy $\dot{\rho}/\Mp < \sigma < 1$
(see Ref.~\cite{IRmulti}, in which we
considered a multi-field model of inflation, decomposing the
fluctuations into the curvature perturbation and the iso-curvature
perturbations. The same
discussion for the iso-curvature perturbation 
applies to the test field discussed here). 
For instance, as the decohered $n$-point functions of $\gR$, 
we proposed to compute
\begin{eqnarray}
  \frac{\langle {\cal P}_\alpha  \gR(t_{\rm obs},\, \bm{x}_1) \cdots \gR(t_{\rm obs},\, \bm{x}_n)
   \rangle}{\langle {\cal P}_\alpha \rangle}\,.
\end{eqnarray}
The introduction of ${\cal P}_\alpha$ removes 
the contamination from the wave packets (parallel worlds) 
that are not correlated with our wave packet at observation.
In Ref.~\cite{IRmulti}, we showed that after the introduction of the
operator ${\cal P}_\alpha$, the superH modes  
$k \alt e^{\rho(t)} \dot{\rho}(t)$ of the iso-curvature
fields are suppressed. 
Although the IRdiv and IRsec can be removed by this prescription, 
the SG for the iso-curvature perturbation 
are still left as an open question. 

The stochastic approach discussed briefly in
Sec.~\ref{SSSec:stochastic} assumes the decoherence 
when the scale exceeds the Hubble scale 
because of the large squeezing of the quantum 
state~\cite{S, SY, Polarski:1995jg,Nakao:1988yi, Nambu:1988je}. 
The decoherence process has been discussed by means of 
the coarse-graining of some degrees of freedom identified as 
environment. As a result, the reduced density matrix evolves from 
the initial pure state to a mixed state. 
(For instance, see Refs.~\cite{Kiefer:2006je, Morikawa:1989xz, Morikawa:1987ci,Tanaka:1997iy} and also
Refs.~\cite{Hu:2008rga, Roura:2007jj, Urakawa:2007dm}.) 
This process is interpreted as the transition from the
initial coherent superposition of many different worlds to the final 
statistical ensemble of them. 
We believe that the stochastic approach
will provide a good approximation to the description of the decohered
fluctuation~\cite{Finelli:2008zg}. 
However, we also think that the stochastic approach is not sufficient 
to discuss the IR regularity issue, because the
quantum nature of fluctuations of long wavelength modes
is omitted by its assumption from the beginning. 

\section{The graviton loops}  \label{Sec:GW}
We have discussed the IR issues of the scalar perturbation so far,
neglecting the tensor perturbation. In this section, we briefly discuss
the IR issues related to the graviton and overview the recent progress.

\subsection{The IR divergence and the secular growth from the graviton loops}
The quadratic action for the tensor perturbation $\delta
\gamma_{ij}$, 
which describes the evolution of the interaction picture field 
$\delta \gamma_{I\, ij}$, is given by 
\begin{eqnarray}
  S_{0, {\rm GW}} = \frac{\Mp^2}{8} \int \dd t\, \dd^3 \bm{x}\, e^{3\rho}
   \left[ \delta \dot{\gamma}_{I\,j}^i \delta \dot{\gamma}_{I\,i}^j -
    e^{-2\rho} \partial^l \delta \gamma^i_{I\,j} \partial_l \delta
    \gamma^j_{I\,i} \right]\,,
\end{eqnarray}
and the equation of motion is given by
\begin{eqnarray}
  \left[ \partial_t^2 + 3 \dot{\rho} \partial_t - e^{-2\rho} \partial^2
  \right] \delta \gamma_{I\, ij} = 0\,.
\end{eqnarray}
We quantize $\delta \gamma_{I\,ij}$ as  
\begin{eqnarray}
  & \delta {\gamma}^i_{I\,j}(x) = \sum_{\lambda=\pm} \int \frac{\dd^3
 \bm{k}}{(2\pi)^{3/2}} \delta \gamma^{(\lambda)}_k(t)
 e^{(\lambda)i}\!_j(\bm{k}) e^{i \sbm{k} \cdot \sbm{x}}
 a_{\sbm{k}}^{(\lambda)} + ({\rm h.c.})\,, \label{Exp:gI}
\end{eqnarray}
where $\lambda$ is the helicity of the tensor perturbation,
$e^{(\lambda)}_{ij}$ are the transverse and traceless 
polarization tensors normalized as 
$e^{(\lambda)i}\!_j(\bm{k}) e^{(\lambda')j}\!_i(\bm{k}) 
=\delta_{\lambda \lambda'}$, and
$a_{\sbm{k}}^{(\lambda)}$ are the annihilation operators which satisfy
\begin{eqnarray}
  \left[ a_{\sbm{k}}^{(\lambda)},\, a_{\sbm{p}}^{(\lambda')\dagger}
  \right] = \delta_{\lambda \lambda'} \delta^{(3)} (\bm{k} - \bm{p})\,.
\end{eqnarray}
Since the equation for $\delta \gamma_k^{(\lambda)}$ is identical to the one 
for a massless scalar field, 
the graviton field in the adiabatic vacuum (see Sec.~\ref{SSec:free}) 
has the almost scale-invariant spectrum in the IR limit as
\begin{eqnarray}
  P_{\rm GW}(k) = 2\, |\delta \gamma_k(t)|^2 = \frac{4}{k^3} 
\left( \dot{\rho}(t_k) \over \Mp \right)^2 \left[1
 +  {\cal O} \left( (k \eta)^2 \right) \right]\,.
\end{eqnarray}
Since the spectrum is
isotropic, the amplitude of $\delta \gamma_{k}$ does not depend on the
helicity and hence we doubled the amplitude. 
Using Eq.~(\ref{Exp:gI}), the variance of the graviton (in the
coincidence limit) is given by
\begin{eqnarray}
 & \langle \delta \gamma_{I\,ij} (x) \delta \gamma_{I\,kl} (x)  \rangle \cr
 &  = \frac{1}{20 \pi^2} \left( \delta_{ik} \delta_{jl} + \delta_{il} \delta_{jk} -
   \frac{2}{3}\delta_{ij} \delta_{kl} \right) \int \frac{\dd k}{k} k^3
 P_{\rm GW}(k) \,.  \label{Exp:spectrumGW} 
\end{eqnarray}
We can see that similarly
to the curvature perturbation, the superH modes in
Eq.~(\ref{Exp:spectrumGW}) yield the IRdiv and IRsec, respectively. 
As in the case of the curvature perturbation $\zeta$,
one possible way to prove the IR regularity 
and the absence of the SG might be showing that
the $n$-point functions are perturbatively expanded with respected 
to $\delta \gamma_{I\, ij}$ associated with IR suppressing
operators.

Tsamis and Woodard claimed that the logarithmic SG due to the graviton
loops can lead to the screening of the cosmological constant in 
Ref.~\cite{TW962}. More recently, Kitamoto
and Kitazawa claimed that the SG from the graviton loops can screen the gauge
coupling as well in Refs.~\cite{Kitamoto:2012vj, Kitamoto:2013rea}. A
related issue is discussed for the U(1) gauge field in
Refs.~\cite{Leonard:2012fs, Leonard:2013xsa}. If the SG due to the graviton loops is really physical, it will provide an
interesting phenomenological impact. However, we should also keep in
mind the subtlety in the interpretation of the calculated results~\cite{GT07}. 
It was shown that the spatially averaged Hubble expansion computed 
by Tsamis and Woodard is not 
invariant under the change of the time slicing and hence the observed
screening effect suffers from the gauge artifact~\cite{Unruh, GT07}.
Focusing on the fact that a conformally coupled scalar field which
interacts with a source which is homogeneously distributed in the spacetime
measures the Hubble expansion rate $\dot{\rho}$ as 
$\dot{\rho}^2(t) \simeq R/12$ where $R$ is the four dimensional 
Ricci scalar, the authors of Ref.~\cite{GT07} computed the 
expectation value of the smeared Ricci scalar in a local region as 
a local counterpart of the Hubble expansion rate. It was shown 
that the smeared Ricci scalar is time independent at least if an 
appropriate UV renormalization is assumed.
This example tells us that once gravitational perturbations are concerned 
computing observable quantities unaffected by the gauge artifact
is quite important~\cite{GT07} (see also Ref.~\cite{ReplyTW} and
Refs.~\cite{Marozzi:2012tp, Marozzi:2013uva}). 
This is in harmony with our claim regarding the IR issues of the curvature perturbation~\cite{IRgauge_L, IRgauge}.

\subsection{Fixing the residual gauge degrees of freedom}
The study of graviton loops is still in progress and we need more
elaborated discussions to provide a conclusive argument. However, we 
think that an important clue to solve this problem 
is in that the homogeneous mode of the graviton $\delta \gamma_{ij}$ 
is also indistinguishable from the residual \gauge degrees of 
freedom~\cite{IRgauge_L, IRgauge}. Among the residual \gauge
transformations, discussed in Sec.~\ref{SSec:RDF}, we focus on 
\begin{eqnarray}
   & x^i \to  e^{-s(t)} \left[e^{-S(t)/2}  \right]^i\!_j\, x^j \label{Exp:GT}
\end{eqnarray}
where $S_{ij}(t)$ is a time-dependent traceless tensor. 
As for the curvature perturbation, computing an invariant quantity under
the dilatation transformation parametrized by $s(t)$
was a key to show the regularity of the correlation functions. 
Intriguingly, 
at the linear level tensor perturbation is 
shifted as 
\begin{eqnarray}
 \delta\gamma_{ij}(x) \to \delta\gamma_{ij}(x)  - S_{ij}(t)\,, \label{Eq:shiftg}
\end{eqnarray} 
analogously to $\zeta$. 
Although the non-linear extension of 
the above transformation is more complicated than in the case of 
$\zeta$,
this observation suggests that analogous proof 
of the IR regularity may work for graviton loops. 

The relation between the IRdiv due to graviton loops
and the homogeneous shift (\ref{Eq:shiftg}) has been pointed out several
times. 
Gerstenlauer {\it et al}.~\cite{GHT11} and
Giddings and Sloth~\cite{GS10} showed that the leading IRdiv of
the graviton loops can be understood as the change of the spatial
coordinates in the form (\ref{Exp:GT}) with $s=0$ due to the
accumulated effect of IR gravitons. 
In Refs.~\cite{IRgauge_L, IRgauge}, 
we found that the graviton one-loop in the calculation of 
$\langle \gR \gR \rangle$ 
becomes regular without restricting the initial vacuum states. 
In this analysis, however, we heuristically adopted 
a particular solution of the Heisenberg equation. 
In Ref.~\cite{next}, we provide a comprehensive study on 
the regularity of graviton loops, focusing on genuinely \gauge invariant
quantities. 

The gravitons can exist also
in the exact de Sitter background unlike the curvature
perturbation. 
In the exact de Sitter background, 
the IR regularity might be rephrased as 
the existence of a regular quantum state that respects the 
de Sitter invariance.
If the two-point function 
in the Euclidean vacuum for free graviton field is regular, 
it will be extended to the higher order perturbation 
straightforwardly without suffering from the IRdiv 
by performing the vertex integrations 
on the Euclidean sphere that is the analytic continuation 
of the de Sitter space. 
Even at the linear level, however, the regularity of the 
graviton two-point function is still under debate. 
In Ref.~\cite{Higuchi:2011vw}, Higuchi {\it et al}. claimed the
existence of a regular two-point function, 
while Miao {\it et al}. objected against it in
Ref.~\cite{Miao:2011ng}. 
This issue was discussed also in Refs.~\cite{Fewster:2012bj,
Higuchi:2012vy, Morrison:2013rqa}. At the moment we are writing this
review article, there is no consensus regarding this issue.

\section{Concluding remarks and future issues}  \label{Sec:summary}
We summarized the issues regarding the loop corrections of the
three different types of perturbation 
in the inflationary universe; the adiabatic
perturbation, the iso-curvature perturbation, and the tensor
perturbation. Irrespective of the type of perturbations, what is
crucially important for ensuring the IR regularity is to remove the
unobservable effects. Namely, unless we concede the influence of the residual \gauge
modes, the adiabatic and tensor perturbations are free from the IR
pathologies. The most intriguing result we have obtained will be that
when we perform the quantization in the global universe, choosing an appropriate quantum
state is required to regularize the IR contributions. 
Fortunately, the IR regularity is guaranteed if we choose 
the ordinary Euclidean vacuum.

In this section, we further address this point asking the question, 
``When we 
require the $n$-point functions to be finite and free from 
the SG, is the Euclidean
vacuum the unique possible quantum state?'' 
If we inquire the regularity of $n$-point
functions on the real time axis all the way back to the distant past, 
we naively expect that the Euclidean vacuum is the unique possibility. 
This is because any excitations are blue-shifted toward the past, 
and hence any small deviation from the Euclidean vacuum 
will be enhanced to an infinite magnitude in the limit. 
On the other hand, if we require the regularity only in the future of a 
given initial hypersurface, we would be able to construct 
a variety of allowed quantum states. (See also the studies by Einhorn and
Larsen in Refs.~\cite{EL02, EL03} and by Marolf {\it et al}. in
Ref.~\cite{Marolf:2012kh}.) 

The residual \gauge degrees of freedom can also affect the notion of 
the tree-level non-Gaussianity. In Ref.~\cite{IRNG}, 
 we re-defined the primordial non-Gaussianity in single
 field models, requesting the genuinely \gauge invariance. 
 We studied the local bi-spectrum, taking the squeezed limit, where one of $\bm{k}_i$ is sent to
 0. Namely, we revisited the so-called 
 consistency relation, which relates the local non-Gaussianity with the power
 spectrum. As is pointed out
 by Maldacena in Ref.~\cite{Maldacena2002}, the leading contribution in
 the consistency relation stems from the effect of the IR mode
 $\bm{k}_1$, which shifts $\bm{k}_j$ with $j=2,3$ as 
$\bm{k}_j \to e^{-\zeta_{\sbm{k}_1}} \bm{k}_j$. As is expected from the
fact that this dilatation transformation is one of the residual \gauge
transformations, the leading contribution in the consistency relation
does not appear when we evaluate the genuinely \gauge invariant three
point function. This analysis is extended to multi-field models of inflation in Ref.~\cite{IRgauge_multi}. 
A related issue has been
 studied recently by Creminelli {\it et al}.~\cite{Creminelli:2011sq} and 
by Pajer {\it et al}.~\cite{Pajer:2013ana}.  These studies indicate that the \gauge degrees of freedom 
should be carefully treated to provide a theoretical prediction 
to compare with observations even at the tree level. 

\ack
This work is supported by the Grant-in-Aid for the Global COE Program
"The Next Generation of Physics, Spun from Universality and Emergence"
from the Ministry of Education, Culture, Sports, Science and Technology
(MEXT) of Japan. T.~T. is supported by Monbukagakusho Grant-in-Aid for
Scientific Research Nos.~24103006, 24103001, 21244033, and
21111006. Y.~U. is supported by the JSPS under Contact No.~21244033, MEC
FPA2010-20807-C02-02, AGAUR 2009-SGR-168 and CPAN CSD2007-00042
Consolider-Ingenio 2010.
Y.~U. would like to thank the organizers and the participants of the
workshop "Physics of de Sitter Spacetime" in the Max-Planck Institute for
Gravitational Physics, during which the IR issues were intensively
discussed. \\

\end{document}